\documentclass[conference]{IEEEtran}

\ifCLASSINFOpdf
\else
\fi
\usepackage{fontawesome}
\usepackage{tikz}

\usetikzlibrary{shapes,arrows,calc,backgrounds,external}
\usepackage[]{units}
\usepackage{balance}
\usepackage{ifthen}
\usepackage{scalerel}
\usepackage{float}

\usepackage{subcaption}
\usepackage{amssymb}
\usepackage{amsmath}
\usepackage{amsfonts}
\usepackage[sort]{cite}
\usepackage{color}
\usepackage{booktabs}
\usepackage{graphics}
\usepackage{graphicx}

\usepackage{acronym}
\usepackage[lined,algonl,ruled]{algorithm2e}
\usepackage{hyperref}

\hypersetup{
 bookmarksopen=true,
 bookmarksnumbered=true,
 colorlinks=false,
 citecolor=blue,
 linkcolor=red,
 linktocpage=true,
 linkbordercolor={1 1 1},
 hypertexnames=false,
 plainpages=true,
 pdfsubject={},
 pdfkeywords={},
 pdftitle={},
 pdfauthor={}
}

\usepackage[normalem]{ulem}
\usepackage{multirow}

\newcommand{\vect}[1]{\mathbf{#1}}

\newlength{\tmpbighat}
\newcommand{\bighat}[1]{
 \setlength{\tmpbighat}{\unitlength}
 \settowidth{\unitlength}{\mbox{$#1$}}
 \ifdim \unitlength < 22pt
  \widehat{#1}
 \else
  \stackrel{\bighatpic}{#1}
 \fi
 \setlength{\unitlength}{\tmpbighat}
}
\newcommand{\bighatpic}{\begin{picture}(1,0.1)(0,0)
 \put(0.05,0){\line(6,1){0.45}}
 \put(0.95,0){\line(-6,1){0.45}}
 \end{picture}}
\newcommand{\RNum}[1]{\uppercase\expandafter{\romannumeral #1\relax}}

\begin{document}

%
\title{peerRTF: Robust MVDR Beamforming \\Using Graph Convolutional Network}

\author{
Daniel Levi, Amit~Sofer and  Sharon~Gannot~\IEEEmembership{Fellow,~IEEE}
\thanks{The authors are with Bar-Ilan University, Israel.  e-mail: \texttt{\{daniel.levi1,amit.sofer,sharon.gannot\}@biu.ac.il}. The work was partially supported by grant \#3-16416 from the Ministry of Science \& Technology, Israel, and from the European Union’s Horizon 2020 Research and Innovation Programme, Grant Agreement No.~871245. Daniel Levi and Amit Sofer equally contributed to the paper.
Project Page: \url{https://peerrtf.github.io/} (including audio demonstration and link to code).
}
}
\markboth{IEEE/ACM Transactions on Audio, Speech, and Language Processing,~Vol.~x, No.~y, Aug.~2024}
{Levi, Sofer and Gannot: {peerRTF: Robust MVDR Beamforming Using Graph Convolutional Network}}

%
%

%

\acrodef{RTF}{relative transfer function}
\acrodef{ML}{manifold learning}
\acrodef{ATF}{acoustic transfer function}
\acrodef{AIR}{acoustic impulse response}
\acrodef{STFT}{short-time Fourier transform}
\acrodef{RT}{reverberation time}
\acrodef{GEVD}{generalized eigenvalue decomposition}
\acrodef{OOG}{out of grid}
\acrodef{MVDR}{minimum variance distortionless response}
\acrodef{SNR}{signal-to-noise ratio}
\acrodef{PSD}{power spectral density}
\acrodef{EVD}{eigenvalue decomposition}
\acrodef{SVD}{singular value decomposition}
\acrodef{STOI}{short-time objective intelligibility}
\acrodef{ESTOI}{extended short-time objective intelligibility}
\acrodef{GCN}{graph convolutional network}
\acrodef{GDL}{geometric deep learning}
\acrodef{GNN}{graph neural network}
\acrodef{CNN}{convolutional neural network}
\acrodef{MLP}{multi-layer perceptron}
\acrodef{KNN}[$\mathcal{K}$NN]{$\mathcal{K}$ nearest neighbors}
\acrodef{ELU}{exponential linear unit}
\acrodef{DOA}{direction of arrival}
\acrodef{LS}{least squares}
\acrodef{VAE}{variational autoencoder}
\acrodef{AE}{autoencoder}
\acrodef{ReLU}{rectified linear unit}
\acrodef{tanh}{hyperbolic tangent}
\acrodef{FC}{fully-connected}
\acrodef{NPM}{normalized projection misalignment}
\acrodef{SBF}{signal blocking factor}
\acrodef{SI-SDR}{scale-invariant source-to-distortion ratio}
\acrodef{DNSMOS}{deep noise suppression mean opinion score}
\acrodef{VAD}{voice activity detection}
\acrodef{DNN}{deep neural network}
\acrodef{ReIR}{relative impulse response}
\acrodef{RIR}{room impulse response}
\acrodef{TDOA}{time difference of arrival}

\acrodef{MP}{manifold projection}
\acrodef{iFFT}{inverse fast Fourier transform}
\acrodef{FFT}{fast Fourier transform}
\acrodef{CoG}{center of grid}



\maketitle



%
\IEEEpeerreviewmaketitle

\begin{abstract}
Accurate and reliable identification of the \acp{RTF} between  
microphones with respect to a desired source is an essential component in the design of microphone array beamformers, specifically when applying the \ac{MVDR} criterion. 
Since an accurate estimation of the \ac{RTF} in a noisy and reverberant environment is a cumbersome task, we aim at leveraging prior knowledge of the acoustic enclosure to robustify the \acp{RTF} estimation by learning the \ac{RTF} manifold.
In this paper, we present a novel robust \ac{RTF} identification method, tested and trained using both real recordings and simulated scenarios, which relies on learning the \ac{RTF} manifold using a \ac{GCN} to infer a robust representation of the \acp{RTF} in a confined area, and consequently enhance the beamformer's performance.

\end{abstract}


%
\IEEEpeerreviewmaketitle

\section{Introduction}
%
%
%

\IEEEPARstart{M}{odern} acoustic beamformers outperform conventional \ac{DOA}-based beamformers due to their ability to consider the entire acoustic propagation path rather than only the direct path. The construction of these beamformers necessitates an estimate of the \acp{AIR} relating the source and the microphones (or their corresponding \acp{ATF}). To alleviate the challenge of blindly estimating the \ac{ATF} it was proposed in \cite{gannot2001signal} to substitute the \acp{ATF} by the \acp{RTF} in the design of the beamformer.
\textcolor{black}{Specifically, the \ac{MVDR} beamformer is a spatial filter designed to minimize the noise power at its output while preserving the desired source without distortion. There is accumulated evidence that justifies the use of the \acp{RTF} as the steering vector for calculating the \ac{MVDR} weights \cite{gannot2001signal,gannot2017consolidated,shmaryahu2022importance}. In our research, we adopt this approach.}

The \ac{RTF} is defined as the ratio between two \acp{ATF}. Specifically, it represents the \ac{ATF} that relates the source to one microphone, normalized by the \ac{ATF} that relates the source to a designated reference microphone.
This definition encapsulates the relative acoustic relation between microphones in an array. It effectively captures the relative differences in how the sound propagates to different microphones, which is crucial for various acoustic signal processing tasks. Various \ac{RTF}-based audio beamformers can be found in the literature, often yielding improved performance compared to \ac{DOA}-based beamformers.
While various algorithms for estimating \acp{RTF} can be found in the literature, such as those proposed in \cite{gannot2001signal, markovich2015performance, li2015estimation, shalvi1996system, koldovsky2015spatial}, they often face degradation in challenging conditions, particularly in low \ac{SNR} environments with high reverberation. 

The literature extensively covers approaches to enhance beamforming robustness, commonly achieved through techniques like beam widening, as discussed in \cite{zheng2004robust, doclo2010acoustic, cox1987robust, li2003robust, li2004doubly,barnov2018spatially}.
 While these methods have shown success, our approach takes a different direction by focusing on improving the estimated \ac{RTF} through leveraging a pre-learned set of \acp{RTF} utilizing a modern manifold technique.

Despite their intricate structure, it is demonstrated in \cite{laufer2015study} that the \acp{RTF} are primarily controlled by a limited set of parameters, such as the size and geometry of the room, the positions of the source and the microphones, and the (frequency-dependent) reflection coefficients of the walls. Consequently, acoustic paths exhibit low-dimensional geometric structures, commonly referred to as manifolds, and can be analyzed using \ac{ML} methods.
In a fixed room with a static microphone array location, the only degree of freedom is the source location, causing the \ac{RTF} to vary only based on the speaker's position. Consequently, \acp{RTF} from different locations lie on a manifold. By assembling a clean set of \acp{RTF} as a training dataset, we can explore the \ac{RTF} manifold and derive a more robust estimate of the \ac{RTF} from noisy recordings.
\textcolor{black}{This result has found various applications in audio, including localization \cite{laufer2016semi, deleforge20122d, laufer2017semi, deleforge2015acoustic}, acoustic scene mapping \cite{Cohen2024unsupervised} and speech enhancement \cite{talmon2013relative,sofer2021robust, brendel2022manifold}. The application of manifold learning to RTF estimation is particularly relevant to our work.}

Several \ac{ML} approaches, such as those proposed by \cite{tenenbaum2000global, roweis2000nonlinear, belkin2001laplacian}, typically follow a standard framework. In this framework, manifold samples are initially represented as a graph. Subsequently, a low-dimensional representation (embedding) of the data is inferred, preserving its structure meaningfully. This representation effectively `flattens' the original non-Euclidean structure of the manifold into an Euclidean space, simplifying subsequent analysis. 

\textcolor{black}{In the context of beamforming, previous efforts to learn the manifold of the \acp{RTF} have employed a graph representation, utilizing the Gaussian heat kernel to determine edge weights \cite{talmon2013relative,sofer2021robust}. Specifically, in \cite{sofer2021robust}, the \ac{RTF} manifold is initially represented by a graph, where the \acp{RTF} serve as graph nodes, and the edges' weights are defined using the heat kernel function. 
A Markov process is established on the graph by constructing a transition matrix representing the manifold diffusion process.
Leveraging spectral graph theory, this approach derives a low-dimensional embedding of the dataset in Euclidean space, where the Euclidean distance between samples reflects the diffusion distance across the high-dimensional manifold surface. The final estimator is created using geometric harmonics \cite{coifman2006geometric}, which extends these low-dimensional embeddings to new data points, enabling supervised \ac{RTF} identification. 
Post-inference, an \ac{MVDR} beamformer is applied using the low-dimensional embedding to accomplish the desired noise reduction task. 
An alternative \ac{RTF}-\ac{ML} approach is proposed in \cite{brendel2022manifold}, employing \ac{VAE} to robustify the \ac{RTF} estimation. The \ac{VAE} is trained in an unsupervised manner using data collected under benign acoustic conditions, enabling it to reconstruct \acp{RTF} within the specified enclosure.
The method introduces a \ac{LS}-based \ac{RTF} estimator that is regularized by the trained \ac{VAE}. This regularization significantly improves the quality of \ac{RTF} estimates compared to traditional \ac{VAE}-based denoising methods. In this way, a hybrid model is devised, combining classic \ac{RTF} estimation with the capabilities of the trained \ac{VAE}. The robust \ac{RTF} estimate can be applied in further processing.
}

In recent years, \ac{GDL}, a term describing techniques that extend deep neural models to non-Euclidean inputs like graphs and manifolds, has seen significant application in classification, segmentation, clustering, and recommendation tasks. Its adoption is more prevalent in fields like social sciences (e.g., analyzing social networks using graphs), chemistry (where molecules can be represented as graphs), biology (where bio-molecular interactions form graph structures), 3D point cloud \ac{ML}, computer vision, and others.
Those methods usually focus on classification, segmentation, clustering, and recommendation tasks but not on regression tasks. \Ac{GNN}, a specific type of \ac{GDL}, specializes in learning representations from graph-structured data by effectively propagating information between interconnected nodes.
A particular type of \ac{GNN} is the \ac{GCN} which is based
on the principles of learning through shared-weights, similar to \acp{CNN} \cite{kipf2022semi, schlichtkrull2018modeling, zhang2019graph, velivckovic2018graph, wang2019dynamic, zhou2020graph}. \Acp{GCN} effectively leverage graph structures by performing convolution operations over the nodes and edges, allowing them to capture the local neighborhood information and aggregate features from adjacent nodes. This approach enables \acp{GCN} to learn meaningful representations of graph-structured data.


 Recent advances demonstrate that \acp{GNN}  naturally emerges in \ac{ML} \cite{bronstein2017geometric}. Inspired by these trends, we aim to substitute the traditional \ac{ML} techniques with methods relying on \ac{GNN}, particularly on \ac{GCN}. The conventional \ac{ML} techniques involve flattening the non-Euclidean manifold into an Euclidean space.
We will harness the power of \ac{GCN} to learn the high dimensional \ac{RTF} manifold and to infer a robust estimator of a \ac{RTF} from noisy measurements thereof by leveraging the graph representation of the manifold.

\textcolor{black}{While other approaches \cite{ito2016complex,erdogan2016improved,heymann2016neural,ochiai2020beamtasnet,drude2019unsupervised} (and the \ac{DNN}-\ac{MVDR} variant of \cite{saijo2022spatial}) employ spectral masking to facilitate the estimation of the beamformer's building blocks, our method takes a different approach by leveraging spatial information from neighboring positions to robustify the steering vector of the beamformer in reverberant environments. This spatial perspective offers a complementary way to enhance beamforming performance. These alternative spectral and spatial approaches can both enhance the accuracy of the beamformer's steering vector estimation. While combining them could yield even better results, such a combination is beyond the scope of this paper.}

Our contribution is \textcolor{black}{threefold}: 1)  a novel robust \ac{RTF} estimation algorithm that infers the \ac{RTF} manifold using a \ac{GCN} and leverages it to robustify the \ac{RTF} estimation; \textcolor{black}{2) a multi-view perspective of \acp{GCN}, effectively combining multiple graphs - a reminiscent of the multiple-manifold learning approach \cite{laufer2017semi}; and 3)} a comprehensive assessment of the proposed scheme and its performance advantages as compared with competing methods in various \ac{SNR} levels, \textcolor{black}{noise types,} and \textcolor{black}{real-world and simulated acoustic responses.} 
\textcolor{black}{To our knowledge, this is the first attempt to enhance \ac{RTF} estimation using \acp{GCN}. We believe this contribution can pave the way for more extensive use of \acp{GNN} in audio processing—a paradigm yet to be explored in the field.
}

\textcolor{black}{The remainder of this paper is organized as follows. 
In Sec.~\ref{Problem Formulation}, we formalize the problem and present the notation used throughout the paper.
Section ~\ref{Proposed Framework} introduces our proposed approach.
Section~\ref{RTFs based MVDR Beamformer} explains a general robust beamforming approach, which includes the vanilla \ac{RTF} estimation and \ac{RTF}-based beamforming.
Section~\ref{Our method}  elaborates on our approach, in particular, the creation of the graph data, the architecture of the \ac{GCN}, and the objective functions.
Section~\ref{Experimental Setup} describes the experimental setup using a dataset of recorded \acp{RIR} and presents the experimental results together with an elaborated comparison with other competing methods. Section~\ref{Additional study} extends the experimental study to lower grid resolution using simulated data and different types of additive noise. The fundamentals of \ac{GCN} are discussed in Appendix~\ref{sec:GCN}. Section~\ref{Conclusion} concludes the paper.}

\section{Background and Problem Setup}
\label{Problem Formulation}
\subsection{Problem Formulation} 
An $M$-microphone array is positioned in a reverberant enclosure. We assume that the desired source location is confined to a known region. Examples of such environments include conference rooms, where the microphone array is placed at a fixed location on the table, and speakers occupy designated positions around it. Similarly, in office setups, the microphone array is fixed on the desk or computer screen, with the speaker typically seated behind the desk. In a car, the microphone array is positioned at a fixed location at the visor, while the speaker occupies one of the seats.

Let $r_m(t), m = 0, \ldots, M-1$, denote the measured signal at the $m$th microphone:
\begin{equation}
\label{eq:Time_dom_mod}
    r_m (t) = \{s * a_m\} (t) + v_m(t),
\end{equation}
with $s(t)$ representing the desired speech signal, and $v_m(t)$ the contribution of all noise sources as captured by the $m$th microphone, ${a}_m(t)$ stands for the \ac{AIR} from the source to the $m$th microphone at time $t$, and $*$ denotes the convolution operator.
In scenarios where the speaker remains static, the \ac{AIR} remains constant over time.
The time-domain convolution in \eqref{eq:Time_dom_mod} can be approximated by multiplication in the \ac{STFT} domain. All $M$ equations can then be written in a single vector form as:
\begin{equation}
\label{eq:STFT_dom_mod}
\mathbf{r} (l,k) = s(l,k) \mathbf{a} (k) + \mathbf{v} (l,k).
\end{equation}
Here, $l$ and $k$ represent the time-frame and frequency-bin indexes, respectively, with $l \in \{0,\ldots,L-1\}$ and $k \in \{0,\ldots,K-1\}$. The vector $\mathbf{a} (k) = [a_0 (k), \ldots ,a_{M-1} (k)]^\top$, comprises all \acp{ATF} from the source to the microphone array.
We define $a_{\textrm{ref}}(k)$ as the component of the vector $\mathbf{a}(k)$ that corresponds to the reference microphone. Equation \eqref{eq:STFT_dom_mod} can also be reformulated as a function of $\tilde{s} (l,k) = s(l,k) a_{\textrm{ref}}(k)$, representing the source signal as captured by the reference microphone:
\begin{equation}
\label{eq:STFT_dom_mod_rtf}
\mathbf{r} (l,k) = \tilde{s} (l,k) \mathbf{h} (k) + \mathbf{v} (l,k),
\end{equation}
where $\mathbf{h}(k)$ is the vector of \acp{RTF}:
\begin{equation}
\mathbf{h}(k) \triangleq \frac{\mathbf{a}(k)}{a_{\textrm{ref}}(k)}.
\end{equation}

\subsection{Nomenclature}
\textcolor{black}{This section introduces the key notations and methods used throughout the paper.
}
\textcolor{black}{As will be described throughout the paper, we will work with both the frequency-domain \acp{RTF} $h_m(k), k=0,1,\ldots,K-1; m=0,\ldots,M-1$, and their time-domain counterparts, the \ac{ReIR}, $\bar{h}_m(n), m=1,\ldots,M-1; n=-n_{\textrm{non-causal}},\ldots,n_{\textrm{causal}}$. The \acp{ReIR} are truncated to obtain smoothing in the frequency domain. 
We define the following vectors:
\begin{itemize}
\item $\bar{\mathbf{h}}_{\textrm{alg}}^m=\textrm{vec}_n\{{\bar{h}}_{\textrm{alg}}^m(n)\}$: Time-domain \ac{ReIR} vector concatenating all taps from $-n_{\textrm{non-causal}}$ to $n_{\textrm{causal}}$ for each microphone $m=1,\ldots,M-1$ (excluding the reference microphone).
\item $\mathbf{h}_{\textrm{alg}}(k)=\textrm{vec}_m\{{h}_{\textrm{alg}}^m(k)\}$: Concatenation of all \acp{RTF} across microphones and for all frequency bins $k$.\footnote{We include the reference microphone \ac{RTF} (which always equals 1) in the concatenated vector to obtain the full $M$-dimensional vector, required for the steering vector of the frequency-domain beamformer.}
\end{itemize}
The subscript $\textrm{alg}$ denotes the estimation method, where $\textrm{alg} \in \{\textrm{gevd}, \textrm{oracle}, \textrm{gcn}, \textrm{mp}, \textrm{vae} , \textrm{cog} , \textrm{meanGrid}\}$, defined as follows:
\begin{itemize}
\item \textrm{gevd}: Vanilla \ac{RTF} estimated by the \ac{GEVD} procedure, followed by \ac{iFFT}, truncation, and \ac{FFT}, as explained in Sec.~\ref{GEVD}.
\item \textrm{oracle}: Clean \ac{RTF} estimated in noiseless environments, followed by \ac{iFFT}, truncation, and \ac{FFT}, as explained in Sec.~\ref{oracle estimation}.
\item \textrm{gcn}: \ac{RTF} obtained by applying the \ac{GCN} (our proposed method) explained in Sec.~\ref{Our method}.
\item \textrm{mp}: The \ac{MP}-based baseline method \cite{sofer2021robust}, explained in Sec.~\ref{Experimental Setup}.
\item \textrm{vae}: The \ac{VAE}-based baseline method \cite{brendel2022manifold}, explained in Sec.~\ref{Experimental Setup}.
\item \textrm{cog}: Uses the \ac{RTF} corresponding to the center position of the measurement cube (center of grid).
\item \textrm{meanGrid}: Computes the average of all training oracle \acp{RTF}, providing a baseline that leverages the entire training set.
\end{itemize}
}

\section{Proposed Framework}
\label{Proposed Framework}
\textcolor{black}{
\ac{RTF}-based \ac{MVDR} beamformers rely on accurately estimated \acp{RTF} to achieve high directivity and minimal distortion of the desired source signal \cite{gannot2001signal,gannot2004analysis,shmaryahu2022importance}. In this work, we leverage spatial information to improve noisy \ac{RTF} estimates. Our framework employs \acp{GCN} to infer an \ac{RTF} (or the respective time-domain \ac{ReIR}) manifold from clean \acp{RTF}. The set of clean \ac{ReIR} are used to construct a graph whose nodes correspond to clean features and whose edges encode their interrelationships. By using the smoothness of the \acp{ReIR}) on the inferred manifold, we obtain a robust representation that serves as input to the \ac{MVDR} beamformer.
The overall architecture of the proposed scheme is depicted in Fig.~\ref{fig:Block_Diagranm}. The experimental setup involves a microphone array placed in an acoustic enclosure. We assume that a set of oracle \acp{ReIR}, measured from a (not necessarily regular) grid of potential source locations within the enclosure, is available.}
\begin{figure*}[ht!]
\centering
\includegraphics[width=0.9\linewidth]{figs/block_diagram_multi_.png}
\captionsetup{justification=centering,margin=0.5cm}
\caption{Block diagram of the proposed robust \ac{RTF}-based beamforming system. The process consists of four main stages: (1) Initial \ac{ReIR} estimation ($\bar{\textbf{h}}_\textrm{gevd}^m$) from noisy signals using \ac{GEVD}, \ac{iFFT}, and truncation, (2) Enhancement of these \acp{ReIR} using the \ac{GCN} architecture and oracle \acp{ReIR} to obtain robust estimates ($\bar{\textbf{h}}_\textrm{gcn}^m$), (3) Transformation to the frequency domain via \ac{FFT} and concatenation across microphones to form the \ac{RTF} vector, and (4) Application of the \ac{MVDR} beamformer using the enhanced \ac{RTF} for final signal estimation.}
\label{fig:Block_Diagranm}
\end{figure*}
\textcolor{black}{The framework consists of several key steps:
1) The \acp{ReIR}, denoted as $\bar{\textbf{h}}_\textrm{gevd}^m$, are estimated from the noisy input signals using \ac{GEVD}, \ac{iFFT}, and truncation; 2) These noisy \acp{ReIR} are enhanced by leveraging oracle \acp{ReIR} from the same acoustic environment through the \ac{GCN} architecture, resulting in robust \acp{ReIR}, denoted as $\bar{\textbf{h}}_\textrm{gcn}^m$;
3) The robustified \acp{ReIR} are transformed back to the frequency domain using \ac{FFT}, followed by concatenation across microphones to form the \ac{RTF} vector; and, finally, the enhanced \ac{RTF} vector is utilized to construct the \ac{MVDR} beamformer, which is applied to the noisy input signals to estimate the desired source signal.
}


\textcolor{black}{
If a set of oracle \acp{ReIR} is available, this approach increases the robustness of \ac{RTF} estimation under noisy and reverberant conditions, thereby significantly improving the \ac{MVDR} beamformer's performance. As discussed, there are scenarios where such oracle \ac{ReIR} measurements can be obtained. This involves playing a sufficiently exciting signal (e.g., pink noise) from multiple positions within the enclosure, ensuring no background noise is present, and then using standard system identification methods. Although this process can be cumbersome, it does not require a precise grid or known source positions.
}

\textcolor{black}{
The following sections provide the theoretical foundations, detailed implementation, and experimental validation of our method, compared against traditional \ac{GEVD} and a state-of-the-art \ac{MP}-based approach.}

\section{RTF-based MVDR Beamformers}
\label{RTFs based MVDR Beamformer}
This section overviews \ac{RTF}-based beamforming for speech enhancement using microphone arrays. The \ac{MVDR} beamformer serves as the backbone algorithm throughout the paper. We first describe the \ac{GEVD}-based approach for estimating \acp{RTF} in noisy conditions, followed by additional estimation procedures. Our main contribution of the paper, namely the robust \ac{RTF} estimation method based on \ac{GCN}, will be detailed in Sec.~\ref{Our method}. 


\subsection{GEVD-Based RTF Estimation - A Concise Overview}
\label{GEVD}
In \cite{markovich2009GEVD,gannot2015cwVScs}, it was demonstrated that the \ac{RTF} could be estimated through the \ac{GEVD} of the spatial correlation matrices of the noisy signal segments $\boldsymbol{\Phi}_{rr,\ell} (k)$\footnote{In the more general form, it can be time-varying, but here we assume that the \ac{RTF} is time-invariant, and can therefore be estimated by averaging over all active-speech time segments.} and of the noise-only signal segments $\boldsymbol{\Phi}_{vv,\ell}(k)$. The latter is estimated from noise-only segments assumed to be available. Here, $\ell$ represents the source position index\textcolor{black}{, which can be associated with the position of the oracle \ac{ReIR} in the training phase or a noisy \ac{ReIR} in either the training or test phases.} The \ac{RTF} is determined by solving
\begin{equation}
\label{eq:gevddef}
\boldsymbol{\Phi}_{rr,\ell} (k) \boldsymbol{\varphi}_{\ell} (k) = 
\mu_{\ell}(k) \boldsymbol{\Phi}_{vv,\ell} (k) \boldsymbol{\varphi}_{\ell} (k).
\end{equation}
Using $\boldsymbol{\varphi}_{\ell} (k)$, the generalized eigenvector corresponding to the largest generalized eigenvalue $\mu_{\ell}(k)$, we obtain the vector of \acp{RTF} 
using the following normalization:
\begin{equation}
\label{eq:gevd}
{\tilde{\mathbf{h}}}_{\textrm{gevd},\ell} (k) = \frac{\boldsymbol{\Phi}_{vv,\ell} (k) \boldsymbol{\varphi}_{\ell} (k)}{\left(\boldsymbol{\Phi}_{vv,\ell} (k) \boldsymbol{\varphi}_{\ell} (k)\right)_{\textrm{ref}}}.
\end{equation}
For each microphone $m$, we obtain the corresponding time-domain representation of the \ac{RTF}, which we denote as \ac{ReIR}, by applying \ac{iFFT} to $\textrm{vec}_k\{{\tilde{h}}^m_{\textrm{gevd},\ell}(k)\}$, a concatenated vector across all frequency bins $k$. The \ac{ReIR} exhibits a distinct pattern characterized by a prominent peak around zero and a rapid decay on both sides. This characteristic allows us to simplify the estimation process by truncating the \ac{ReIR} around its central region, thereby reducing the number of data points that need to be estimated. This truncation also results in a smoothness of the \ac{RTF} in the frequency domain. Specifically, we truncate the \ac{ReIR} to $n=-n_{\textrm{non-causal}},\ldots,n_{\textrm{causal}}-1$ taps. We denote the truncated \ac{ReIR} as $\bar{h}_{\textrm{gevd},\ell}^m(n)$, for $m=1,\ldots,M-1$ and $n=-n_{\textrm{non-causal}},\ldots,n_{\textrm{causal}}-1$, and concatenate all taps to form the vector $\bar{\mathbf{h}}_{\textrm{gevd},\ell}^m$. We omit the index $m=0$ as it corresponds to the reference microphone, and its \ac{ReIR} is a trivial impulse.

\subsection{The Oracle RTF Estimation}
\label{oracle estimation}
The clean, oracle \ac{RTF} estimation procedure provides reference measurements for our method. These oracle \acp{ReIR} will later serve as features of the graph vertices in our graph construction.

We obtain the oracle \ac{RTF} by applying the \ac{GEVD} procedure to noiseless training recordings. In the absence of noise, $\boldsymbol{\Phi}_{vv,\ell}(k)$ in \eqref{eq:gevddef} is substituted by an identity matrix, simplifying \eqref{eq:gevddef} to an \ac{EVD} problem. 
Similar to the noisy case, we apply an \ac{iFFT} followed by a truncation operation to the tap range $n=-n_{\textrm{non-causal}},\ldots,n_{\textrm{causal}}-1$. For $m=1,\ldots,M-1$, we denote the truncated \acp{ReIR} as $\bar{h}_{\textrm{oracle},\ell}^m(n)$ where $n=-n_{\textrm{non-causal}},\ldots,n_{\textrm{causal}}-1$, and concatenate all taps to form the vector $\bar{\mathbf{h}}_{\textrm{oracle},\ell}^m$. 

\subsection{Training and Testing Notation}
\textcolor{black}{The proposed \ac{GCN}-based beamformer necessitates a training dataset. For implementing and evaluating this robust beamformer, we distinguish between training and testing scenarios. Let $\alpha$ denote training position index and $\beta$ denote testing position index.}

\textcolor{black}{For the training set, define $\bar{\mathbf{h}}_{\textrm{oracle},\alpha}^m$ for $\alpha = 1, \ldots, {N_{\textrm{train}}}$ and  $m = 1 ,\ldots, M-1$ as the \ac{ReIR} associated with the $\alpha$-th training position and the $m$-th microphone. The set of all \ac{ReIR} training points associated with the $m$-th microphone is denoted as $\bar{\mathcal{H}}^m = \{\bar{\mathbf{h}}_{\textrm{oracle},\alpha}^m\}_{\alpha=1}^{N_{\textrm{train}}}$}.

\textcolor{black}{For the test set, let $\mathbf{h}_{\textrm{alg},\beta}(k)$ represent the \ac{RTF} vector at the $\beta$-th test position, where $\beta = 1, \ldots, N_{\textrm{test}}$, and $\textrm{alg}$ represents the estimation method as defined previously, and will be elaborated in the following sections.}

\subsection{The MVDR Beamformer}

Let $\mathbf{h}_{\textrm{alg},\beta}(k)$ represent the \ac{RTF} vector
from our test data at a specific position, estimated by one of the designated algorithms.
Define $\boldsymbol{\Phi}_{vv,\beta}(k)$ as the $M \times M$ spatial \ac{PSD} matrix at the $k$-th frequency bin of the noise signals. It is assumed that noise-only segments are available and can be identified, e.g., by applying a \ac{VAD}.

The \ac{MVDR} beamformer is a spatial filter designed to minimize the noise power at its output while maintaining a distortionless response toward the desired source. Its optimal weights are given by:
\begin{equation}
\label{eq:MVDR}
\mathbf{w}^{\textrm{MVDR}}_{\textrm{alg},\beta}(k) = \frac{\boldsymbol{\Phi}_{vv,\beta}^{-1}(k)\mathbf{h}_{\textrm{alg},\beta}(k)}{\mathbf{h}_{\textrm{alg},\beta}(k)^{\mathsf{H}}\boldsymbol{\Phi}_{vv,\beta}^{-1}(k)\mathbf{h}_{\textrm{alg},\beta}(k)}.
\end{equation}
Following  \cite{gannot2001signal} and subsequent publications, we 
use the \ac{RTF} as the steering vector of the \ac{MVDR} beamformer. 
Multiple studies have shown (see, e.g., \cite{gannot2017consolidated,shmaryahu2022importance}) that this \ac{RTF}-based approach significantly outperforms traditional \ac{DOA}-based beamforming in reverberant environments.

\section{peerRTF: A GCN-based Robust RTF Estimation}
\label{Our method}
This section introduces the proposed robust \ac{RTF} estimation method. We delve into the preprocessing of the data, the construction of a feature vector, and the associated graph data. Finally, we explore the derived \ac{GCN} architecture and our objective functions. \textcolor{black}{A concise summary of the principles of \acp{GCN} and their relation to manifold learning can be found in Appendix~\ref{sec:GCN}.}
Our method is inspired by the manifold-learning approaches presented in \cite{talmon2013relative,sofer2021robust}. In the current contribution, we propose to harness a modern  \ac{GCN}-based \ac{ML} methodology to obtain an accurate and robust estimator of \acp{RTF} in noisy and reverberant environments.
Similar to the previous works, our approach leverages prior knowledge regarding the acoustic environment to project the noisy samples onto the manifold. Given that our data is represented as a graph, we utilize message-passing techniques to achieve this goal.  
\textcolor{black}{Figure~\ref{fig:Robust_RTFs} describes the full architecture. Details are provided in the following subsections.}

\begin{figure*}[htbp]
\centering
\includegraphics[width=\textwidth]{figs/Robust_RTFs_inference_fix.png}
\captionsetup{justification=centering,margin=0.5cm}
\setlength{\belowcaptionskip}{-10pt}
\caption{The robust \ac{RTF} estimator during inference stage. The inputs to the system are clean and noisy \acp{ReIR}. The clean \acp{ReIR} are obtained in the training phase, while the noisy \acp{ReIR} are estimated at the test phase. The graph is constructed by applying the \ac{KNN} procedure. Subsequently, a \ac{GCN} is applied on the graph, resulting in the robustified \acp{ReIR}. Note that there are $M-1$ parallel graphs with sharing weights in the \ac{GCN}.}
\label{fig:Robust_RTFs}
\end{figure*}
\subsection{Graph Representation of ReIRs}

The learning process involves understanding the relations between neighboring entities. In our case, this requires learning the \ac{GNN} weights. Before training, we need to construct the graph, including defining the relations between nodes.
This section describes the feature vectors, graph construction, training, and test procedures.

\subsubsection{Feature Vector}
\begin{figure}[htbp]
\includegraphics[width=0.48\textwidth]{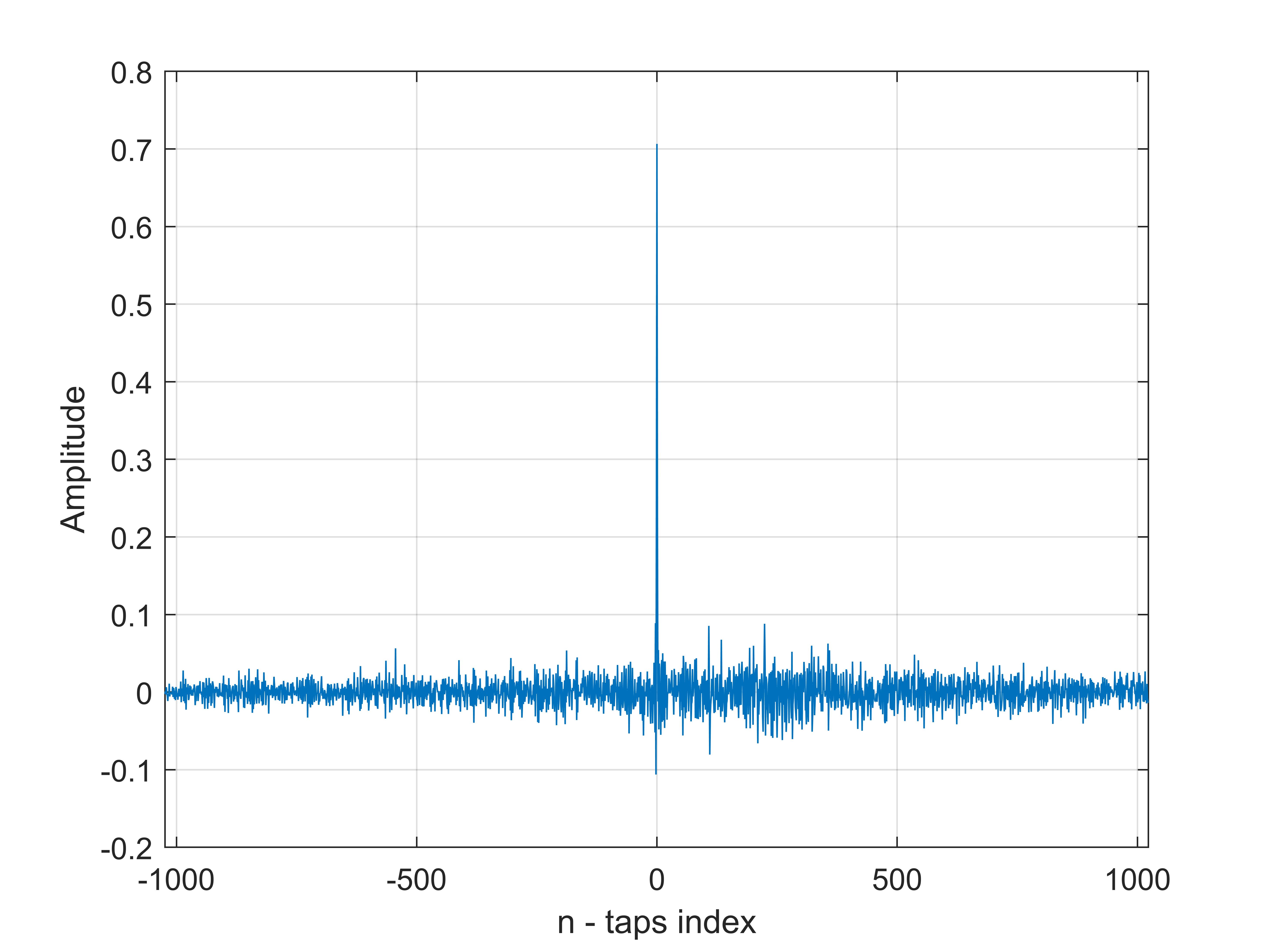}
\setlength{\belowcaptionskip}{3pt}
\caption{Typical \ac{ReIR} corresponding to \acp{RIR} from the MIRaGe database with a reverberation time of $T_{60}=300 \text{ms}$.}
\label{fig:tdrtf}
\end{figure}

While \ac{RTF}-based beamformers are applied in the frequency domain, the corresponding time-domain \ac{ReIR} offers key advantages, mainly due to the rapid decay on both sides of the main peak, as explained above. Additionally, working in the time domain circumvents the need to work with either complex-valued networks or the statistically correlated real and imaginary parts of the frequency-domain representation, thus simplifying the learning process. Figure~\ref{fig:tdrtf} depicts a typical \ac{ReIR} associated with \acp{AIR} from the MIRaGe dataset \cite{mirage_new}, with reverberation time of $T_{60}=300 \text{ms}$.
This example represents a typical \ac{ReIR} associated with one of the grid points. The \ac{RTF} vector is computed using the \ac{GEVD} procedure \eqref{eq:gevd} under noiseless conditions, where an identity matrix substitutes the spatial correlation matrix of the noise. The clean microphone signals are generated by convolving \acp{AIR} from the MIRaGe dataset with a pink noise input signal. We select one of the $M-1$ \acp{RTF} from this grid point, transform it to the time domain, and finally truncate it to obtain the \ac{ReIR}.

For an array of $M$ microphones, each speaker location is associated with $M-1$ \acp{RTF}, as the \ac{RTF} between the reference microphone and itself is trivial. These $M-1$ components are typically estimated independently. The truncated \ac{ReIR} has dimension $d = n_{\textrm{non-causal}} + n_{\textrm{causal}}$, significantly smaller than the full \ac{ReIR}, which enhances learning capabilities. We construct $M-1$ separate graphs, one for each microphone pair, where each room location contributes $M-1$ features of dimension $d$.

\subsubsection{Graph Construction}
%
Building upon these features, we construct a separate graph for each microphone pair,  where each graph comprises  $N_\textrm{train}$ nodes. For each graph, the node features are the oracle \acp{ReIR} set, $\bar{\mathcal{H}}^m = \{\bar{\mathbf{h}}_{\textrm{oracle},\alpha}^m\}_{\alpha=1}^{N_{\textrm{train}}}$.

The graph is constructed by applying a \ac{KNN} procedure, which selects the most similar \acp{ReIR} (in terms of Euclidean distance) from the dataset. This allows us to effectively robustify the \acp{ReIR} for the noisy feature vectors by leveraging information from relevant neighbors. \textcolor{black}{A key advantage of this approach is that it operates directly on the \ac{ReIR} similarities, circumventing the tedious task of explicitly determining position labels, thus making it robust to scenarios where exact source positions might be unknown or imprecise.} Additionally, the use of separate graphs for each microphone pair helps to capture specific relationships and dependencies within the data.

\subsubsection{Training Procedure}

In the training stage, our goal is to learn optimal weights that will enable noisy feature enhancement during testing. Starting with the clean feature sets $\bar{\mathcal{H}}^m$, we iterate through each training position as follows: 1) Exclude a clean feature vector $\bar{\mathbf{h}}_{\textrm{oracle},\alpha_0}^m$ associated with a selected position from all $M-1$ graphs; 2) Incorporate the corresponding noisy feature vectors $\bar{\mathbf{h}}_{\textrm{gevd},\alpha_0}^m$, related to this same position into the graphs using the \ac{KNN} procedure. 

Consequently, each training example thus comprises $N_{\textrm{train}} - 1$ clean feature vectors and one noisy feature vector corresponding to the position that was removed.

\subsubsection{Test Procedure}
In the test phase, we have $N_{\textrm{test}}$ vanilla \acp{ReIR} estimated in noisy scenarios: $\bar{\mathbf{h}}_{\textrm{gevd},\beta}^m$ where $\beta=1,\ldots,{N_{\textrm{test}}}$.
\textcolor{black}{These test samples are processed sequentially. For each microphone pair, the corresponding noisy feature vector from the test sample is added to its respective graph using the \ac{KNN} procedure. After adding a test sample, the respective graph comprises $N_{\textrm{train}} + 1$ nodes, each with a feature vector of dimension $d$.
This process is repeated for all $M-1$ graphs, effectively adding one new node to each graph at each step.}

\subsection{The GCN Architecture}
A key factor in the success of \acp{CNN} is their ability to design and reliably train deep models that extract higher-level features at each layer. This is facilitated by weight sharing, where the same kernel is applied across each channel, enabling efficient feature extraction.

\textcolor{black}{In contrast, training deep \ac{GCN} architectures is more challenging. Several studies have highlighted the limitations of \acp{GCN} when stacked in multiple layers, particularly due to issues like vanishing gradients and over-smoothing \cite{cao2020comprehensive, zhou2020graph,li2019deepgcns, li2018deeper}. Consequently, most state-of-the-art \acp{GCN} typically use no more than four layers to aggregate information from neighbors. This limitation refers specifically to the \ac{GCN} depth—that is, how many neighbor orders are aggregated—rather than the neural network depth within each layer, which typically consists of a simple transformation (e.g., a \ac{FC} layer followed by a non-linear activation). While such shallow \ac{GCN} architectures are often sufficient for tasks like classification, segmentation, clustering, and recommendation, they lack the expressive power needed for more complex tasks, such as regression on high-dimensional data. In our scenario, where nodes represent truncated \ac{ReIR} associated with different room positions, we choose not to aggregate information from second-order neighbors. Instead, we implement a deep network with three layers to ensure sufficient expressive power for regression tasks on a high-dimensional abstract manifold. Drawing inspiration from \cite{wang2019dynamic}, which learns 3D manifolds from point clouds, we consider $\bar{\mathbf{h}}_{\textrm{gevd},i}^m$ as the central ``pixel" and $\bar{\mathbf{h}}_{\textrm{oracle},i(j)}^m, j\in\mathcal{N}(i)$ as the surrounding ``patch". To calculate the contribution of each neighboring node $\bar{\mathbf{h}}_{\textrm{oracle},i(j)}^m$ within each graph, we concatenate the feature vector of the central node $\bar{\mathbf{h}}_{\textrm{gevd},i}^m$ with the feature vector of each neighbor $\bar{\mathbf{h}}_{\textrm{oracle},i(j)}^m$ and pass this concatenated vector through the neural network. The neural network output is then aggregated from all neighbors of $\bar{\mathbf{h}}_{\textrm{gevd},i(j)}^m,j \in\mathcal{N}(i)$. }
When deliberating on selecting an aggregation function, it is essential to consider the essence of our regression task on the manifold. Given that our objective is to predict a continuous value falling within the range of the input values, this criterion guides our choice of aggregation functions. In this context, sum and max are not optimal choices. Instead, we opt for the mean operation, explicitly $\frac{1}{|\mathcal{N}(i)|} \sum_{j\in\mathcal{N}(i)}(\cdot)$.
Figure~\ref{fig:arcit} details the selected architecture.
\usetikzlibrary{positioning}
\def\layersep{1.2cm}
\usetikzlibrary{decorations.pathreplacing}

\tikzset{darkstyle/.style={circle,draw,fill=gray!40,minimum size=1.5em}}
\tikzset{centerstyle/.style={circle, inner sep=0.5pt,draw,fill=blue!40,minimum size=1.5em}}
\tikzset{neighbors/.style={circle, inner sep=0.5pt,draw,fill=teal!40,minimum size=1.5em}}
\begin{figure}[htbp]
\centering
\begin{tikzpicture}[draw=black!50, node distance=\layersep,transform shape]
	
	 \scalebox{0.91}{
	\begin{scope}[xshift=-0.5cm,yshift=3cm,rotate=90,shorten >=1pt,->]

    \tikzstyle{neuron}=[circle,fill=black!25,minimum size=17pt,inner sep=0pt]
    \tikzstyle{input neuron}=[neuron, fill=teal!50];
    \tikzstyle{input neuron left}=[neuron, fill=blue!50];
    \tikzstyle{output neuron}=[neuron, fill=red!50];
    \tikzstyle{hidden neuron}=[neuron, fill=yellow!50];
    \tikzstyle{annot} = [text width=4em, text centered]
    \tikzset{hoz/.style={rotate=-90}}   
    \foreach \name / \y in {1,2}
        \node[input neuron left] (I-\name) at (0,-\y) {};
        
    \foreach \name / \y in {3,4}
        \node[input neuron] (I-\name) at (0,-\y) {};

    \foreach \name / \y in {1,...,5}
        \path[yshift=0.5cm]
            node[hidden neuron] (H2-\name) at (\layersep,-\y cm) {};

    \foreach \name / \y in {1,...,5}
        \path[yshift=0.5cm]
            node[hidden neuron] (H-\name) at (2*\layersep,-\y cm) {};
    
    \node[output neuron, right of=H-3] (0) {};
    \node[output neuron, right of=H-2] (1) {};
    \node[output neuron, right of=H-4] (2) {};

    \foreach \source in {1,...,4}
        \foreach \dest in {1,...,5}
            \path (I-\source) edge (H2-\dest);
            
    \foreach \source in {1,...,5}
        \foreach \dest in {1,...,5}
            \path (H2-\source) edge (H-\dest);

    \foreach \source in {1,...,5}
        \foreach \dest in {0,1,2}
            \path (H-\source) edge (\dest);


    
    \draw [decorate,decoration={brace,amplitude=10pt},xshift=-0.4cm,yshift=-1.7cm, blue!50]
(0,-0.75) --  (0,1) node [midway,xshift=-0.6cm, rotate=-90, black!100] 
{
$\bar{\mathbf{h}}_{\textrm{gevd},i}$};

\draw [decorate,decoration={brace,amplitude=10pt},xshift=4cm,yshift=-1cm]
(0,-0.5) --  (0,-2.5) node [midway, rotate=-90,xshift=0.1cm,yshift=0.5cm, black!100] 
{
$\vect{message}_{i,j}$};

\draw [decorate,decoration={brace,amplitude=10pt},xshift=-0.4cm,yshift=-3.7cm, teal!50]
(0,-0.75) --  (0,1) node [midway,xshift=-0.6cm, rotate=-90, black!100] 
{
$\bar{\mathbf{h}}_{\textrm{oracle},i(j)}$};


\node[anchor=center,rotate=-90] at (-2, -2.5) {(a)};
  
	\end{scope}}
	
	 \scalebox{0.83}{
	    \begin{scope}[xshift=8cm, yshift=5cm]

    \node[centerstyle] (v0) at (0:0) [fill] {$\bar{\mathbf{h}}_{\textrm{gevd},i}$};
	\node[neighbors] (v1) at ( 20:1.95) [fill] {$\bar{\mathbf{h}}_{\textrm{oracle},i(1)}$};
	\node[neighbors] (v2) at ( 72:2) [fill] {$\bar{\mathbf{h}}_{\textrm{oracle},i(2)}$};
	\node[neighbors] (v3) at (2*75:1.9) [fill] {$\bar{\mathbf{h}}_{\textrm{oracle},i(3)}$};
	\node[neighbors] (v4) at (3*70:2) [fill] {$\bar{\mathbf{h}}_{\textrm{oracle},i(4)}$};
	\node[neighbors] (v5) at (4*75:1.6) [fill] {$\bar{\mathbf{h}}_{\textrm{oracle},i(5)}$};
	\node[darkstyle] (v6) at (100:3) [fill] {};
	\node[darkstyle] (v7) at (130:3) [fill] {};
	\node[darkstyle] (v8) at (120:4) [fill] {};
	\node[darkstyle] (v9) at (250:3) [fill] {};
	\draw[<-, very thick] (v0)[purple] -- node[above,color=black] { } (v1);
	\draw[<-, very thick] (v0)[purple] -- node[left,color=black,xshift=0.1cm] { } (v2);
	\draw[<-, very thick] (v0)[purple] -- node[above,color=black,yshift=-0.05cm,xshift=0.1cm] { } (v3);
	\draw[<-, very thick] (v0)[purple] -- node[left,color=black,yshift=0.1cm,xshift=0.1cm] { } (v4);
	\draw[<-, very thick] (v0)[purple] -- node[left,color=black,yshift=-0.06cm,xshift=0.1cm] { } (v5);
	\draw (v7) -- (v3);
    \draw (v8) -- (v7);
    \draw (v6) -- (v3);
    \draw (v6) -- (v2);
    \draw (v6) -- (v8);
    \draw (v9) -- (v4);
    \draw (v2) -- (v7);
    \draw (v9) -- (v5);
    
\node[anchor=center] at (-1, -3.5) {(b)};
	\end{scope}}

\end{tikzpicture}
 \caption{Left: The massage passed from the $j$th neighbor of the $i$th node is calculated by concatenating $\bar{\mathbf{h}}_{\textrm{gevd},i}$ and $\bar{\mathbf{h}}_{\textrm{oracle},i(j)}$ and passing this concatenation through the neural network.
\\
Right: The representation of the $i$th node at the output is calculated by aggregating the messages from all the nodes in $\mathcal{N}(i)$. For each microphone, there is a separate graph, and the neighbors are arbitrarily numbered.
\\
*inspired by \cite{wang2019dynamic}.}
\label{fig:arcit}
\end{figure}
We utilized message passing, one of several commonly used methods in \ac{GNN}. As mentioned, this process involves information exchange between nodes and their neighbors on the graph, enabling them to update their knowledge based on local interactions. Message passing facilitates effective learning and inference in graph-based models.
For our graphs, we have  $\mathcal{K}$ representing the number of neighbors.

Our neural network architecture consists of three \ac{FC} layers, followed by an activation function. The input to the network is a concatenated vector of length $2d$, and the architecture can be represented as follows:
$2d \xrightarrow{} 2d \xrightarrow{} 2d  \Rightarrow d  $.
Here, each $\xrightarrow{}$ represents a single \ac{FC} layer followed by a \ac{ReLU} activation function, while $\Rightarrow$ denotes only an \ac{FC} layer. 

Our \ac{GCN} architecture employs two levels of weight sharing. The first level, a standard convention in \acp{GCN}, involves sharing weights across all node connections within each graph. This allows the network to process nodes uniformly regardless of their position in the graph. The second level, specific to our approach, extends weight sharing across all $M-1$ graphs corresponding to different microphone pairs. This means that a single set of \ac{GCN} parameters is used to simultaneously process all microphone pair graphs.
To evaluate the effectiveness of this approach, we experimented with an alternative configuration. In this alternative, we used $M-1$ individual \acp{GCN}, each dedicated to a specific microphone pair graph, working independently without sharing weights across different graphs. This setup allowed for specialized processing of each microphone pair's data, resulting in a simpler training procedure with more parameters.
However, our experiments showed that this separate \acp{GCN} architecture did not yield any significant performance improvements over the shared-weight approach. Given these results, we opted for the shared-weight architecture across all graphs. This decision offers two key advantages: 1) it significantly reduces the overall model complexity by decreasing the number of parameters, and 2) it provides flexibility, allowing the architecture to adapt easily to varying numbers of microphone pairs, a significant consideration in practical applications.

\textcolor{black}{This flexibility is particularly valuable since the \ac{RTF}, similar to \ac{TDOA}-based steering vectors, is defined between microphone pairs, with performance improving as the number of microphones increases. Our method is versatile and can be applied to any number of microphones and array constellations. Furthermore, even if some microphone pairs are unavailable during inference, the method remains effective by utilizing the available microphones. Since the \ac{GCN} processes each \ac{RTF} separately, the method retains its robustness and effectiveness even when some microphones are missing during inference.}

\subsection{Objective Functions}
\begin{figure}[htbp]
\includegraphics[width=0.48\textwidth]{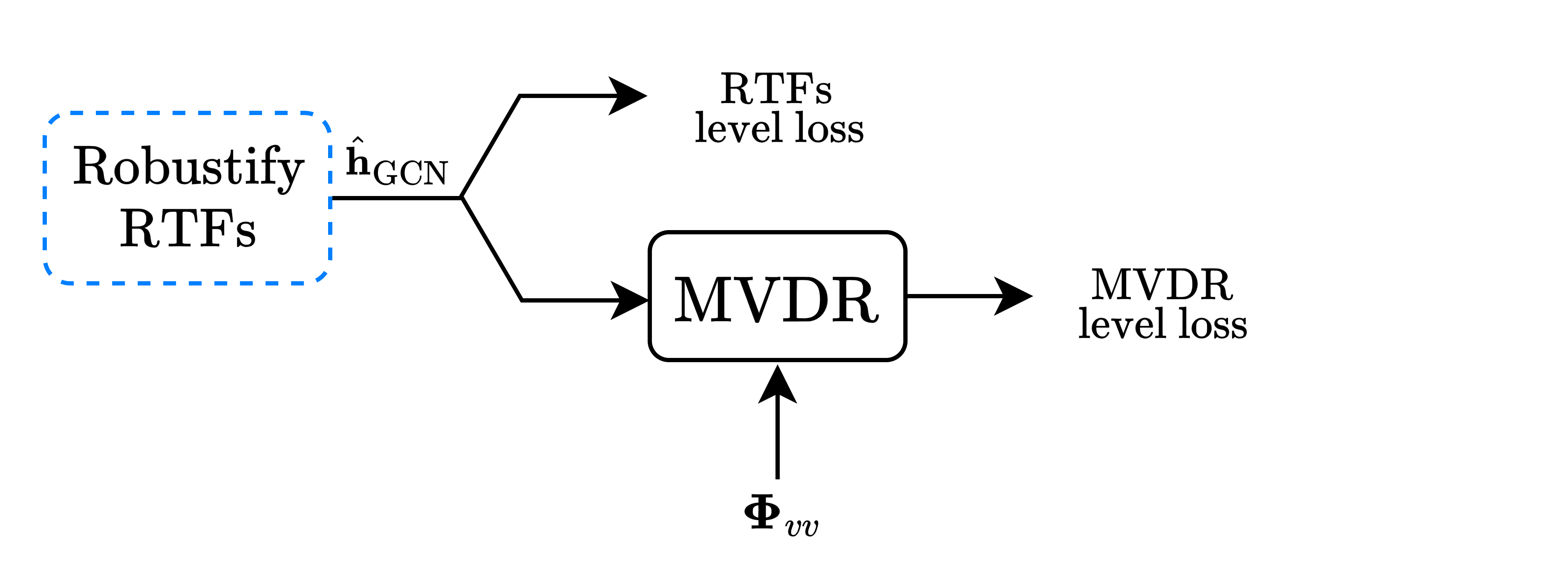}
\centering
\caption{\normalsize Two alternative training objectives.}
\label{fig:train_objectives}
\end{figure}

To efficiently train the model, we examined two alternative objective functions. In the first alternative, we directly optimized the outcome of the \ac{GCN}, namely the \ac{ReIR} estimate. In the second alternative, we optimize the output of the \ac{MVDR} beamformer by adjusting the \ac{RTF} estimate. The two training objectives are schematically depicted in Fig.~\ref{fig:train_objectives} and detailed in the sequel.

\subsubsection{Direct Optimization of the ReIR}

Inspired from~\cite{GannotTalmon}, define the \ac{SBF} as:
\begin{equation}
\textrm{SBF} = \frac{1}{M-1}\sum_{m=0, m\neq \textrm{ref}}^{M-1} 10  \log_{10}\left(\frac{\sum_{t}x_m^2(t)}{\sum_{t}d_m^2(t)}\right)
\end{equation}
where
$$x_m(t)=\{\bar{h}_{\textrm{oracle}}^m \ast \tilde{s}\}(t)$$
and $$d_m(t)= \{\bar{h}_{\textrm{oracle}}^m*\tilde{s}\}(t)-\{\bar{h}_{\textrm{gcn}}^m*\tilde{s}\}(t).$$
Here, $\tilde{s}(t)$ is the reference signal, $\bar{h}_{\textrm{oracle}}^m(t)$ is the oracle \ac{ReIR} corresponding to the $m$th microphone, and $\bar{h}_{\textrm{gcn}}^m(t)$ is the robust \ac{ReIR} of the $m$th microphone. The term $d_m(t)$ is defined as the difference between convolution of $\bar{h}_{\textrm{oracle}}^m(t)$ and $\tilde{s}(t)$ 
with the convolution of $\bar{h}_{\textrm{gcn}}^m(t)$ and $\tilde{s}(t)$.
This function encourages the robust \ac{ReIR} to be as close as possible to the oracle \ac{ReIR}.

\subsubsection{RTF Estimation via Beamformer Output Optimization}

Here, we optimize the \ac{SI-SDR} at the output of the beamformer. The \ac{SI-SDR} is defined as:
\begin{equation}
\label{eq:SISDR}
    \textrm{SI-SDR}\left( \tilde{\vect{s}},\hat{\vect{s}} \right)= 10 \log_{10} \left( \frac{\|{\frac{\langle {\tilde{\vect{s}},\hat{\vect{s}}} \rangle}{\langle {\tilde{\vect{s}},\tilde{\vect{s}}} \rangle} \tilde{\vect{s}}}\|^2}{\|{\frac{\langle {\tilde{\vect{s}},\hat{\vect{s}}} \rangle}{\langle {\tilde{\vect{s}},\tilde{\vect{s}}} \rangle} \tilde{\vect{s}}-\hat{\vect{s}}}\|^2} \right)
\end{equation}
where $\tilde{\vect{s}}$ represents a concatenation of all samples of the reference source, and $\hat{\vect{s}}$ represents the respective vector of all beamformer's output samples.
The \ac{SI-SDR} loss is a metric commonly used to evaluate the quality of source separation or speech enhancement algorithms\cite{le2019sdr}. It measures the enhancement quality between the estimated source signal and the true source signal, considering both the distortion and the interference introduced during the enhancement process.
This loss term aims to bring the beamformer output closer to the clean reference signal. The \ac{RTF} estimate should be adjusted accordingly.
Additionally, we explore an alternative approach by examining the \ac{SI-SDR} compared to the output of the oracle \ac{RTF} beamformer. Here, we compute the \ac{MVDR} weights using the \acp{RTF} estimated under ideal conditions, namely the oracle scenario, and evaluate the resulting \ac{SI-SDR} compared with this signal. This approach aligns more closely with a supervised paradigm, akin to the \ac{RTF} level loss. Importantly, it eliminates the necessity for a clean reference signal in the loss function, addressing a common limitation in scenarios where such a reference signal is unavailable. Still, for this choice, we need the oracle \acp{RTF} to be available, which is another limitation.
We will designate the first version as \ac{SI-SDR}~\RNum{1} and the second as \ac{SI-SDR}~\RNum{2}.

Additionally, we incorporate an implementation of \ac{STOI} as a loss function.\footnote{adopted from \url{https://github.com/mpariente/pytorch_stoi}.} This metric evaluates speech intelligibility and is integrated with \ac{VAD} to focus only on active speech segments.

\textcolor{black}{Algorithm~\ref{alg:algorithm_1} succinctly summarizes the procedural steps for \ac{GCN}-based \ac{RTF} estimation.}

\begin{algorithm}
\caption{Robust \ac{RTF} Estimation Using \ac{GCN}.}
\label{alg:algorithm_1}
\sloppy
\textbf{Training Stage:}
\begin{enumerate}
    \item Build the graphs using \ac{KNN} for each microphone pair using clean \acp{ReIR} $\bar{\mathcal{H}}^m$.
    \item Select one grid position, remove the clean feature vectors ${\mathbf{\bar{h}_{\textrm{oracle},\alpha_0}}}^m$, replace with noisy feature vectors ${\mathbf{\bar{h}_{\textrm{gevd},\alpha_0}}}^m$, and connect to the graphs using \ac{KNN}.
    \item Train \ac{GCN} for robust \ac{ReIR} representation.
    \item[] Repeat $\forall \alpha$, the entire dataset until convergence.
\end{enumerate}

\textbf{Inference Stage:}
\begin{enumerate}
    \item Add a noisy feature vector ${\mathbf{\bar{h}_{\textrm{gevd},\beta_0}}}^m$ to each of the $M-1$ trained graphs using \ac{KNN}.
    \item Process the noisy feature through the trained \ac{GCN} to obtain improved \ac{ReIR} estimates.
    \item[] Repeat $\forall \beta$, test positions.
\end{enumerate}
\end{algorithm}

\section{Experiments with the MIRaGe Dataset} \label{Experimental Setup}
The proposed method is evaluated using 
the MIRaGe dataset \cite{mirage_new}, comprising measured multichannel signals recorded at the Bar-Ilan University acoustic lab. We present a comprehensive evaluation of the proposed \ac{GCN} method through both objective and subjective performance measures and analyze how different graph structures affect the results.

\subsection{Experimental Setup} The MIRaGe database was generated by placing a loudspeaker on a grid of points in a cube-shaped volume with dimensions $46 \times 36 \times 32$~[cm]. The loudspeaker positions were set every $2$~[cm] along the `x' and `y' axes and every $4$~[cm] along the `z' axis, totaling $ 24 \times 19 \times 9 = 4104 $ possible source positions (grid vertices).
In addition, 16 other positions, referred to as \ac{OOG}, were designated as possible locations for noise sources.
The setup was recorded using six static linear microphone arrays, each consisting of $M=5$ microphones with an inter-microphone spacing of $-13[\textrm{cm}], -5[\textrm{cm}], 0, +5[\textrm{cm}], +13[\textrm{cm}]$ relative to the central microphone (the reference microphone). Recordings were made at three different reverberation levels: $100, 300, 600$~ms.

For our experiments, we utilized microphone array \#2, positioned directly in front of the cube at a distance of $2$[m] from its center. The recordings were randomly split into $N_{\textrm{train}}=3500$ training positions, $N_{\textrm{validation}}=100$, and $N_{\textrm{test}}=504$.
We use $2048$ frequency bins, and after the inverse Fourier transform, we truncate the length of the \ac{ReIR} to $n_{\textrm{non-causal}} = 128$ and $n_{\textrm{causal}} = 256$.

For the experimental study, the estimation of the \acp{RTF} involves three steps: 1) The \acp{AIR} from the source position to the microphone arrays are estimated using a \ac{LS} procedure on the recorded chirp signals 2) For clean \acp{RTF} estimation, pink noise signals covering all relevant frequencies, are convolved with the \acp{AIR};\footnote{\textcolor{black}{In real-life scenarios, we may substitute, for practical reasons, the pink noise by ``spontaneous'' speech signals uttered in the environment.}} for estimating the \acp{RTF} from noisy signals, speech signals are convolved with the \acp{AIR} and mixed with pink noise from \ac{OOG} locations 3) The \acp{RTF} are estimated using the \ac{EVD} procedure for clean signals and the \ac{GEVD}-based procedure \eqref{eq:gevd} for noisy utterances.

To construct the training set, we add three independent noise signals—each played from a different position randomly selected from the 16 \ac{OOG} locations—to each of the 3,500 clean training speech signals. The noise signals are mixed with random \ac{SNR} values in the range $[-10, 10]$~dB. This process, using the 16 different \ac{OOG} locations, results in a total of 10,500 samples. The speech signals are sourced from the Librispeech dataset \cite{7178964}.

The network was trained using a linear scheduler with a warmup ratio of 0.1, a learning rate of $1 \times 10^{-4}$, and a dropout rate of 0.5 over 100 epochs. We set $\mathcal{K}=5$ as the \ac{KNN} parameter. We chose \ac{SI-SDR}~\RNum{2} as the objective function for all reverberation times. When comparing different objective functions (detailed in the next section), we observed slight advantages with \ac{SI-SDR}~\RNum{2}, though all objective functions performed well, demonstrating the robustness of our method.
%
The various parameters are listed in Table~\ref{table:Parameters}. 
\begin{table}[htbp]
\setlength{\belowcaptionskip}{-3pt}
\caption{Parameters.}
\begin{center}
\begin{tabular}{@{}llc@{}}
\toprule
Parameter & Description & Value \\
\midrule
$M$ & Number of microphones & 5 \\
$K$ & Number of frequency bins & 2048 \\
\midrule
$n_{\textrm{non-causal}}$  &  Number of taps left of the peak & 128 \\
$n_{\textrm{causal}}$ & Number of taps right of the peak & 256 \\
$\mathcal{K}$ & Number of neighbors in the graph & 5 \\
\bottomrule
\end{tabular}
\end{center}
\label{table:Parameters}
\vspace{-.8cm}
\end{table}

\subsection{Quality Measure}
The results are analyzed using several quality metrics to assess different aspects of the enhanced signal.
The first is the \ac{SNR} at the beamformer's output, calculated as:
\begin{equation}
\textrm{SNR}\left( \hat{\vect{s}},\hat{\vect{v}} \right)= 10 \log_{10} \left( \frac{\left\|{\hat{\vect{s}}}\right\|^2}{\left\|\hat{\vect{v}}\right\|^2} \right).
\end{equation}
Here, $\hat{\vect{s}}$ represents the speech component at the beamformer output, with all samples concatenated into a vector, and $\hat{\vect{v}}$ represents the corresponding noise component.  Since our dataset is simulated and the \ac{MVDR} beamformer is linear, we can apply the beamformer separately to the speech and noise components and obtain $\hat{\vect{s}}$ and $\hat{\vect{v}}$ directly.

Additionally, we assess the signal quality using \ac{STOI} \cite{stoi} for speech intelligibility and \ac{DNSMOS} \cite{reddy2022dnsmos} for overall speech quality. We also examined the \ac{SI-SDR} \eqref{eq:SISDR} in its first variant, comparing the beamformer's output to the reference signal.

\subsection{Baseline Methods}
The proposed \ac{GCN}-based method is compared with six other baselines, all employing the \ac{MVDR} beamformer.  
The first two basic baselines are \ac{RTF}-based \ac{MVDR} beamformers, namely   
%
%
\textbf{the traditional \ac{GEVD} (1)} procedure for \ac{RTF} estimation, using truncated \ac{ReIR}, and \textbf{ the oracle \ac{RTF} (2)} estimated under noise-free conditions, with truncated \ac{ReIR} for fair comparison. 

Another baseline utilizes \textbf{the \ac{MP} learning (3)} method introduced in \cite{sofer2021robust} to robustify \ac{RTF} estimation. This method requires two parameters:  the kernel scale parameter $\epsilon$ and
the number of dominant eigenvalues $\lambda$.  We set $\epsilon = 0.3$ for all reverberation times, while $\lambda$ varies with reverberation level: $\lambda = 12$ for $T_{\textrm{60}} = 100~\text{ms}$, $\lambda = 5$ for $T_{\textrm{60}} = 300~\text{ms}$, and $\lambda = 15$ for $T_{\textrm{60}} = 600~\text{ms}$.

\textcolor{black}{The \textbf{\ac{VAE}-based (4)} baseline approach \cite{brendel2022manifold}, which employes  an unsupervised variational autoencoder trained on clean \acp{RTF} to enhance noisy \ac{RTF} estimates. The \ac{VAE} learns a manifold representation of the \acp{RTF} and uses this learned structure to denoise new estimates. For a fair comparison, we adapt their method to fit our \ac{ReIR} feature vector rather than the original \ac{RTF} estimation, which may have different initial conditions affecting performance. We train the model using our dataset, following the protocol described in the original paper. Furthermore, we adopted the fine-tuning variant, which involves additional training on a noisy training set to improve performance.}

\textcolor{black}{Finally, we compare our method with two additional approaches:
\textbf{\Ac{CoG} (5)} that always uses the \ac{RTF} that corresponds to the center position of the measurement cube, and \textbf{ Mean Grid (6)} approach, which computes the average of all training oracle \acp{RTF}, providing a baseline that leverages the entire training set without considering the noisy signal nor source position.
To ensure a fair comparison, we apply \ac{ReIR} truncation across all methods.
}

\subsection{Results}

\begin{figure*}[htbp]
     \centering
    \begin{subfigure}[b]{0.31\textwidth}
         \centering         \includegraphics[width=\textwidth]{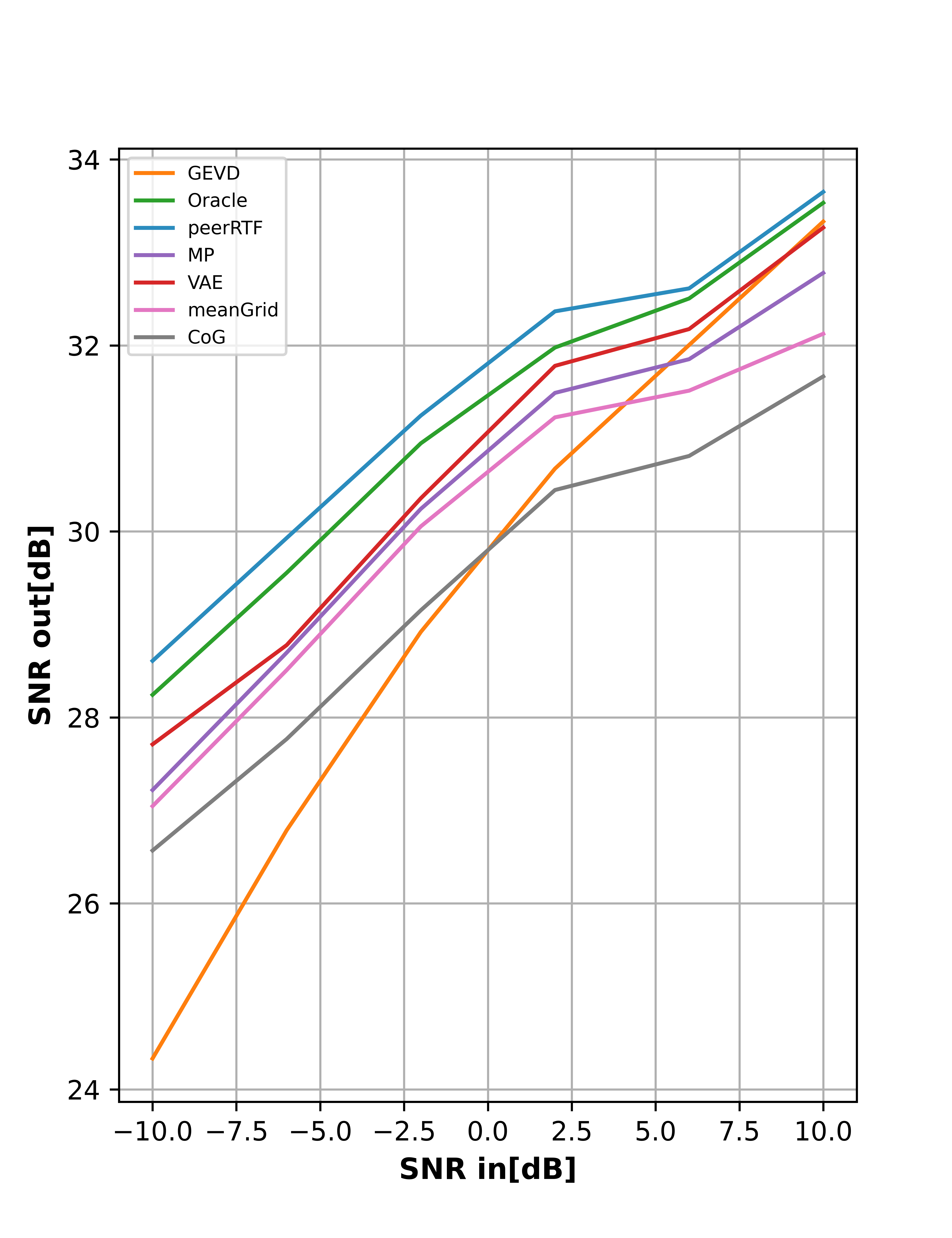}        
    \end{subfigure}
    \begin{subfigure}[b]{0.31\textwidth}
         \centering         \includegraphics[width=\textwidth]{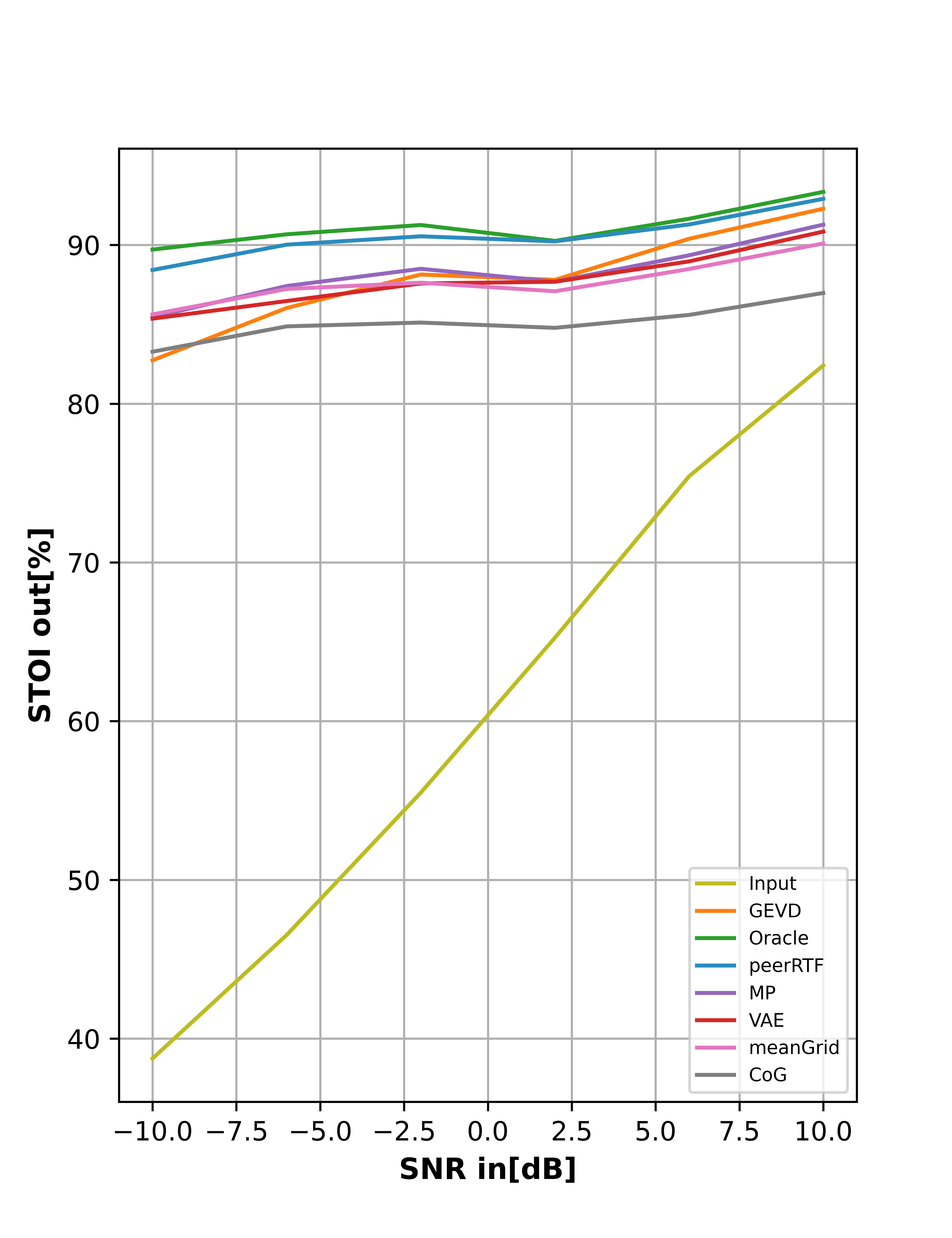}
     \end{subfigure}
    \begin{subfigure}[b]{0.31\textwidth}
         \centering         \includegraphics[width=\textwidth]{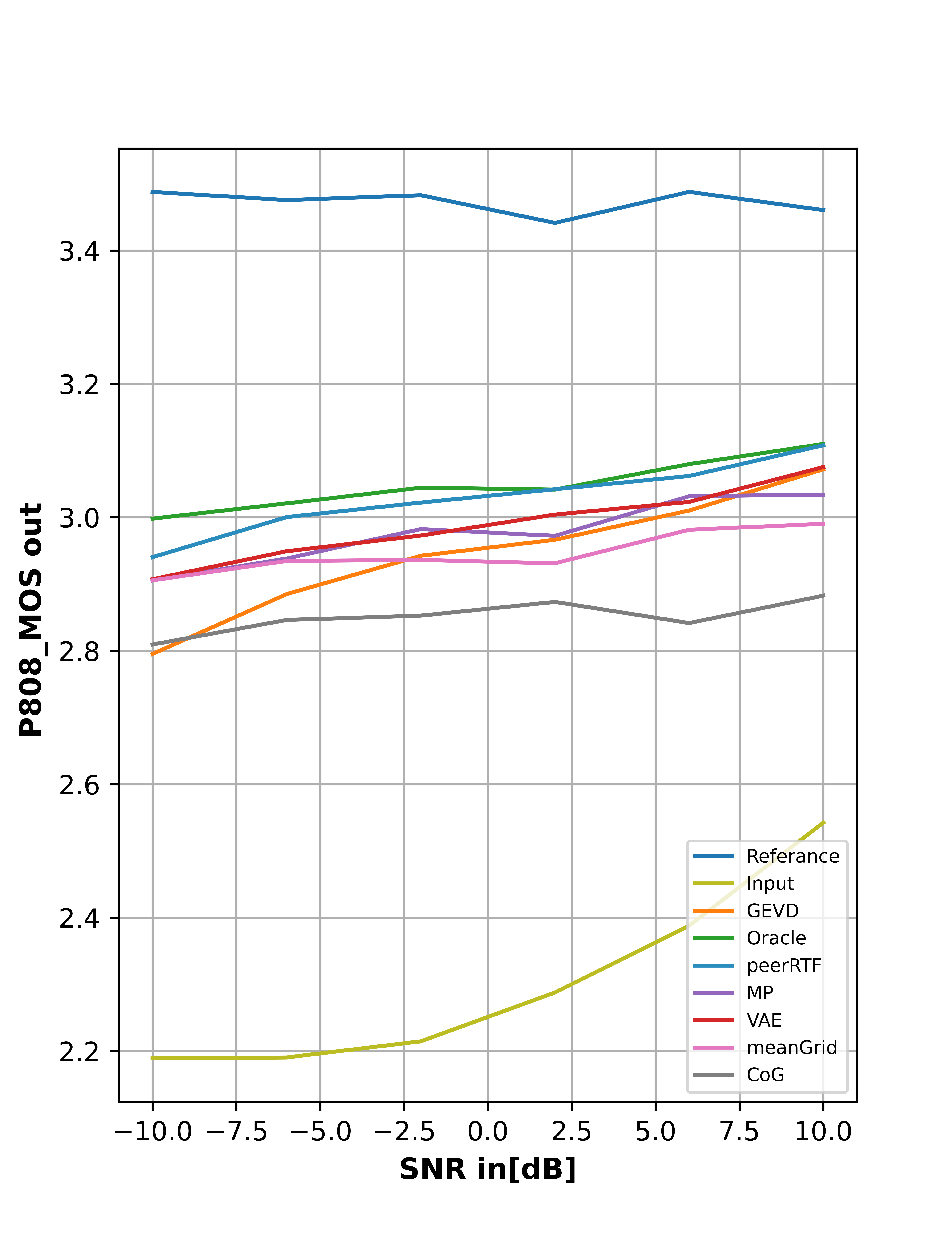}
     \end{subfigure}
\captionsetup{justification=centering,margin=0.5cm}
\caption{$\textrm{SNR}_{\textrm{out}}$[dB] (left), STOI[\%] (middle) and DNSMOS(right) as function of the input \ac{SNR} for MIRaGe dataset, $T_{60}=100$[ms]. Comparison between peerRTF and baseline methods.}

        \label{fig:results100}
\end{figure*}

\begin{figure*}[htbp]
     \centering
    \begin{subfigure}[b]{0.31\textwidth}
         \centering         \includegraphics[width=\textwidth]{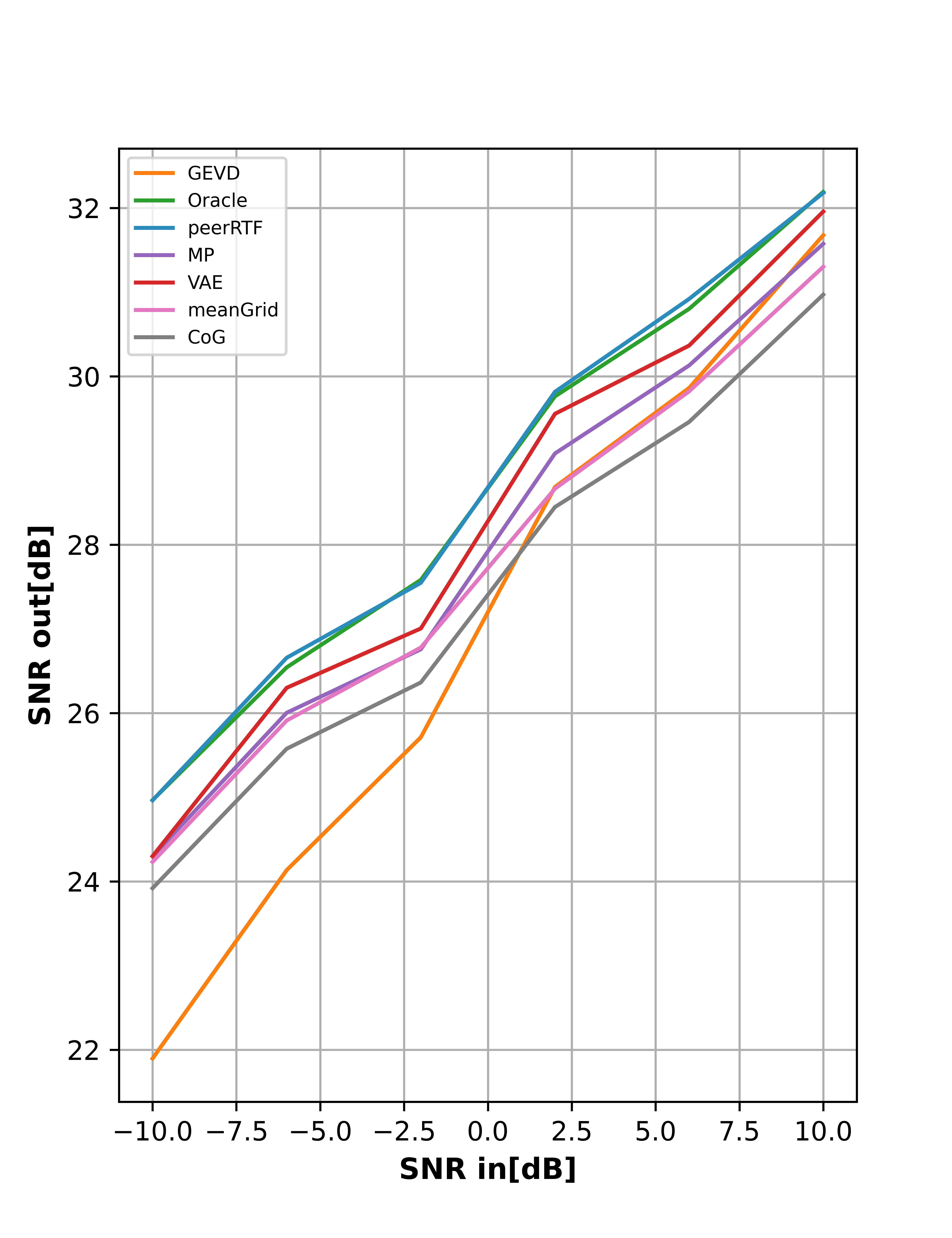}        
    \end{subfigure}
    \begin{subfigure}[b]{0.31\textwidth}
         \centering         \includegraphics[width=\textwidth]{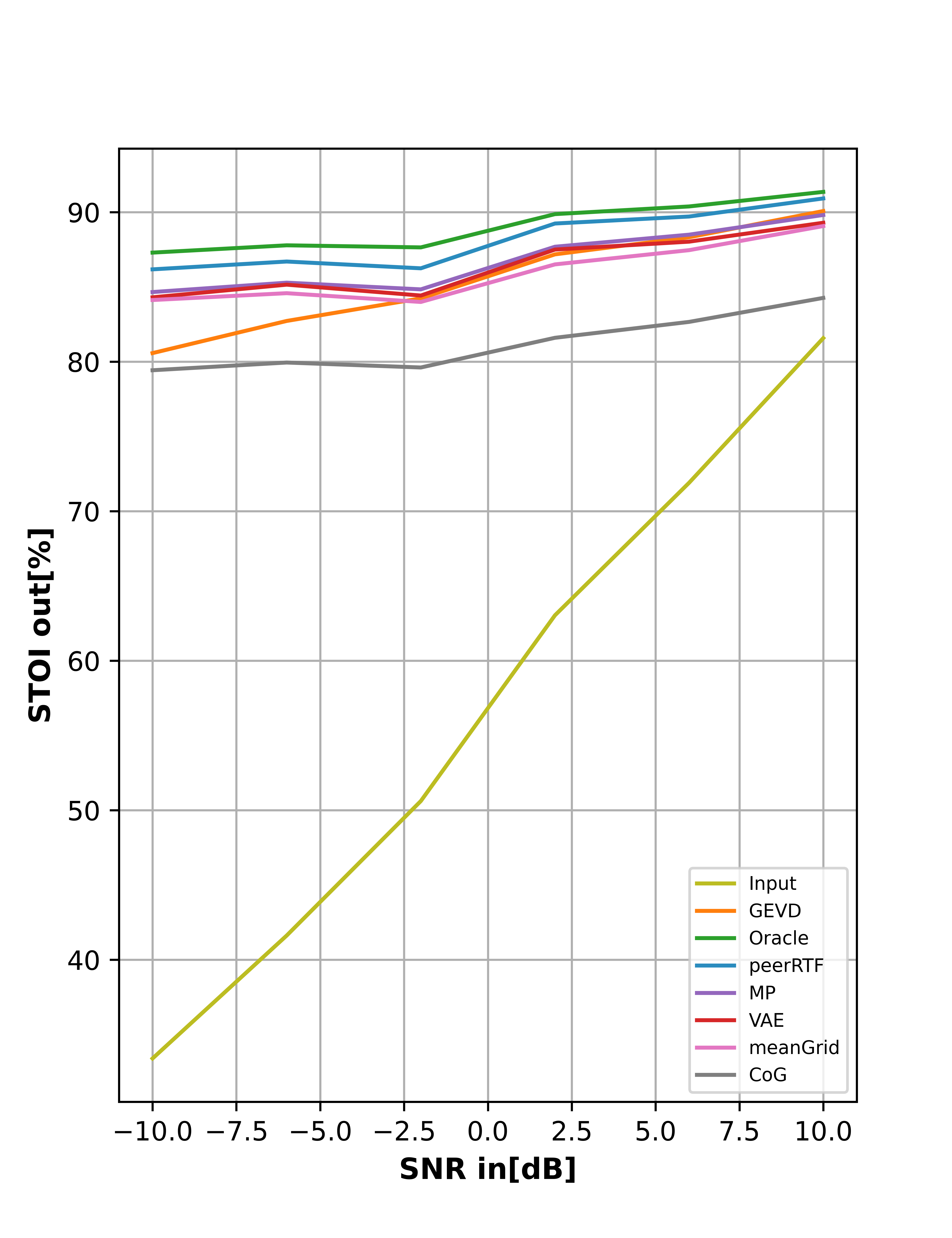}
     \end{subfigure}
    \begin{subfigure}[b]{0.31\textwidth}
         \centering         \includegraphics[width=\textwidth]{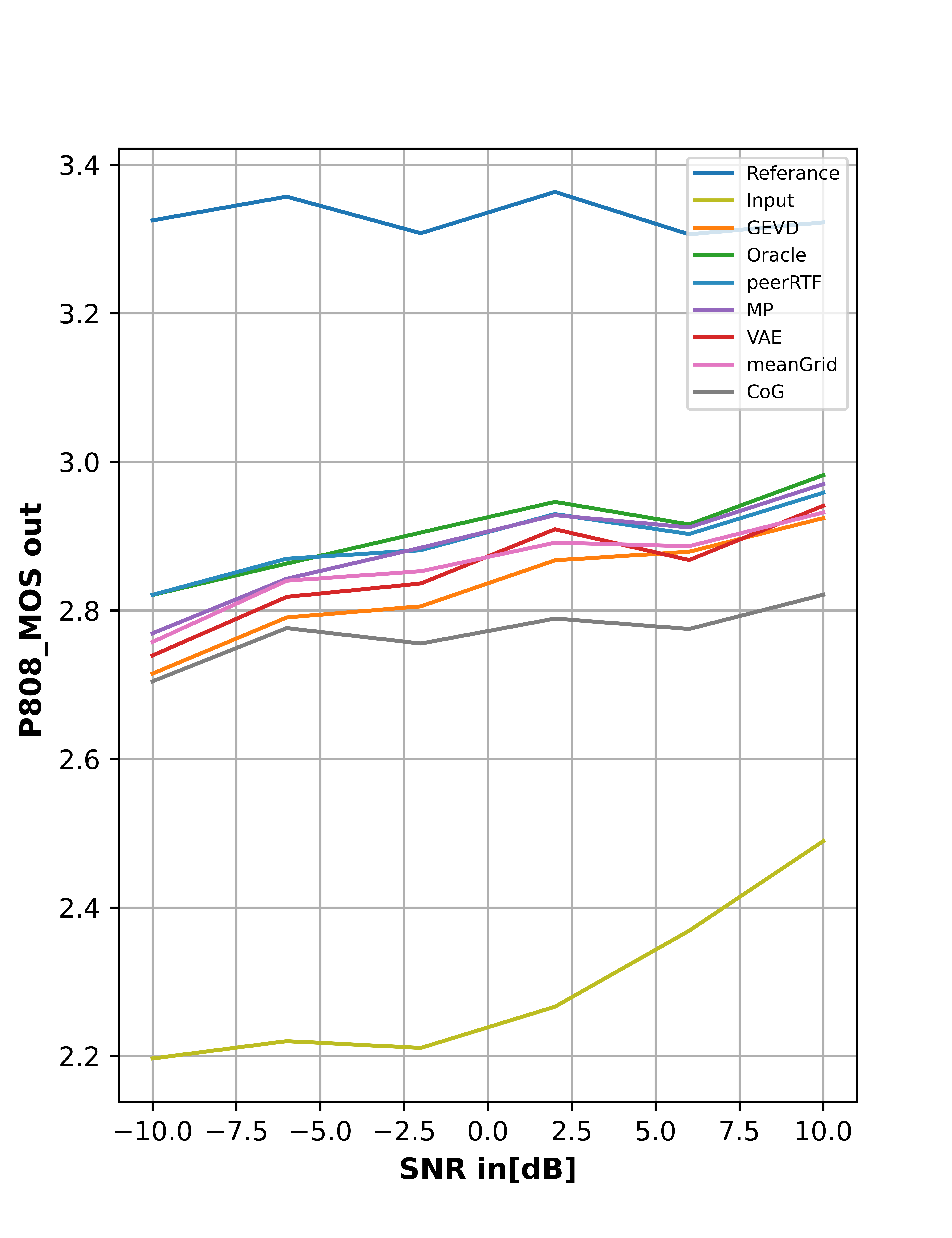}
     \end{subfigure}
\captionsetup{justification=centering,margin=0.5cm}
\caption{$\textrm{SNR}_{\textrm{out}}$[dB] (left), STOI[\%] (middle) and DNSMOS (right) as function of the input \ac{SNR} for MIRaGe dataset, $T_{60}=300$[ms]. Comparison between peerRTF and baseline methods.}
        \label{fig:results300}
\end{figure*}

\begin{figure*}[htbp]
     \centering
    \begin{subfigure}[b]{0.31\textwidth}
         \centering         \includegraphics[width=\textwidth]{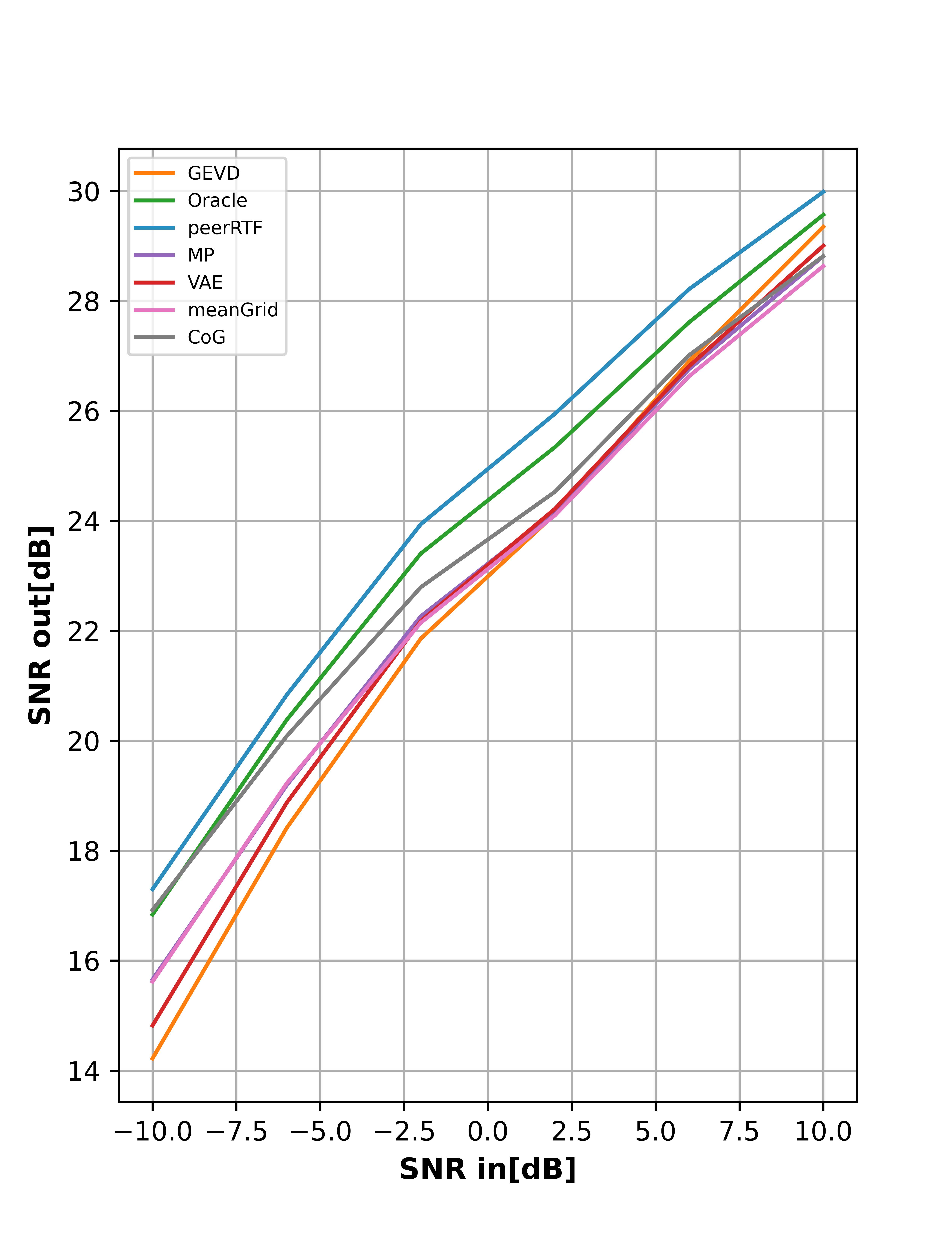}        
    \end{subfigure}
    \begin{subfigure}[b]{0.31\textwidth}
         \centering         \includegraphics[width=\textwidth]{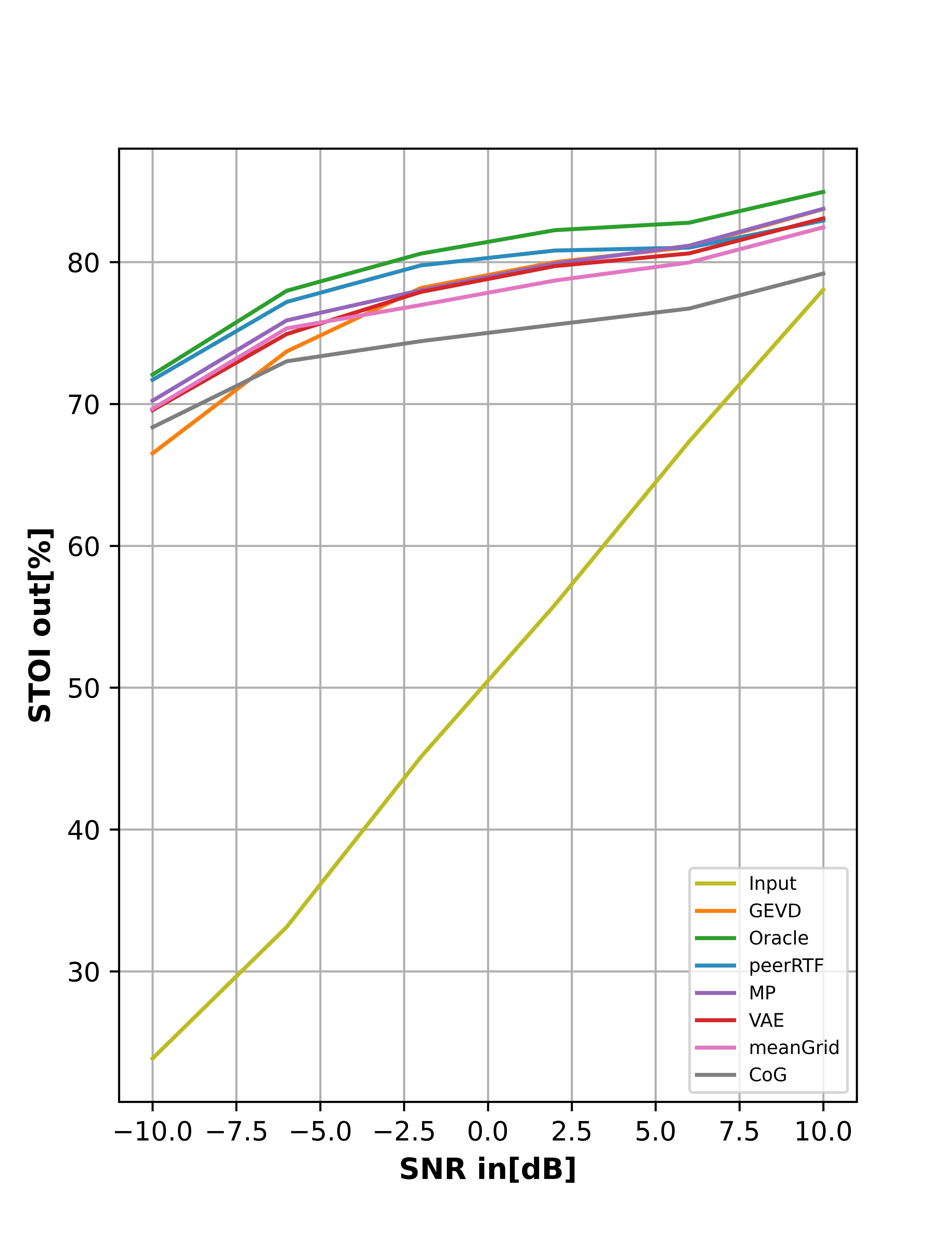}
     \end{subfigure}
    \begin{subfigure}[b]{0.31\textwidth}
         \centering         \includegraphics[width=\textwidth]{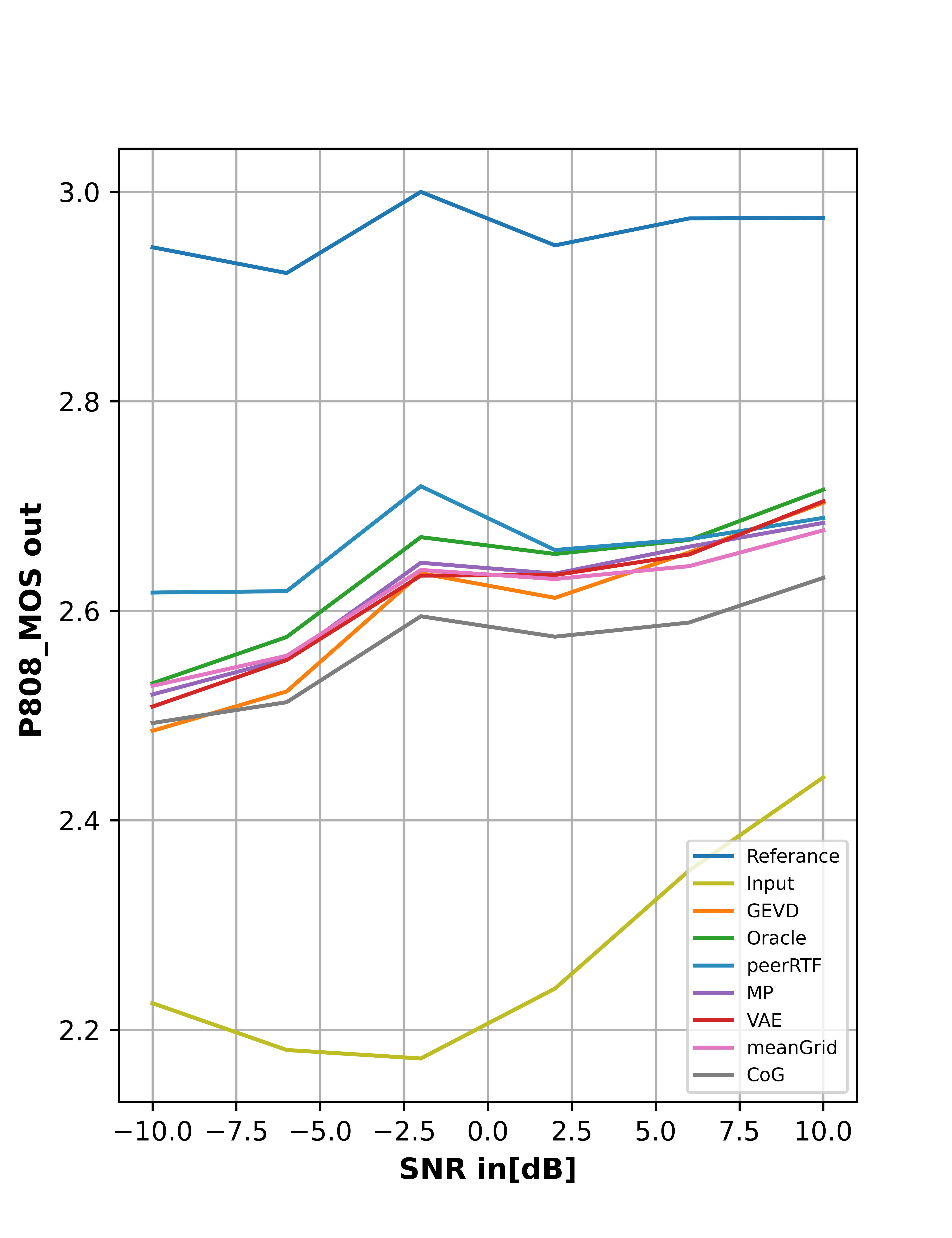}
     \end{subfigure}
\captionsetup{justification=centering,margin=0.5cm}
\caption{$\textrm{SNR}_{\textrm{out}}$[dB] (left), STOI [\%] (middle) and DNSMOS(right) as function of the input \ac{SNR} for MIRaGe dataset, $T_{60}=600$[ms]. Comparison between peerRTF and baseline methods.}

        \label{fig:results600}
\end{figure*}

Figures~\ref{fig:results100},\ref{fig:results300}, and \ref{fig:results600} present the performance comparison between the proposed method (peerRTF) and baseline approaches (\ac{GEVD}, Oracle, \ac{MP}, \ac{VAE}, \ac{CoG}, and Mean Grid) across three reverberation times ($T_{\textrm{60}}= 100, 300, 600~\text{ms}$). For each condition, we evaluate 
$\ac{SNR}_{\textrm{out}}$, \ac{STOI}, and \ac{DNSMOS} as functions of the input \ac{SNR}.
 
\textcolor{black}{The results demonstrate several key findings. First, our proposed method consistently outperforms the vanilla \ac{GEVD}-based beamformer in speech intelligibility across all \ac{SNR} and reverberation levels. When compared to the \ac{MP} and \ac{VAE} beamformer, we observe improvements across most \ac{SNR} levels(especially low \ac{SNR}) and reverberation time. The \ac{CoG} approach occasionally outperforms vanilla \ac{GEVD}, while Mean Grid shows better performance than \ac{GEVD} in most conditions. However, both still show lower performance compared to our method.
The \ac{SNR} at the beamformer output is consistently higher than that of the vanilla \ac{GEVD}, \ac{MP}, \ac{VAE}, \ac{CoG}, and Mean Grid beamformers across all input \ac{SNR} levels and reverberation conditions. Furthermore, our method even outperforms the oracle \ac{RTF} in several \ac{SNR} levels and reverberation times.
}

\begin{figure}[htbp]
     \centering
     \begin{subfigure}[t]{0.23\textwidth}
         \centering
         \includegraphics[width=0.95\textwidth]{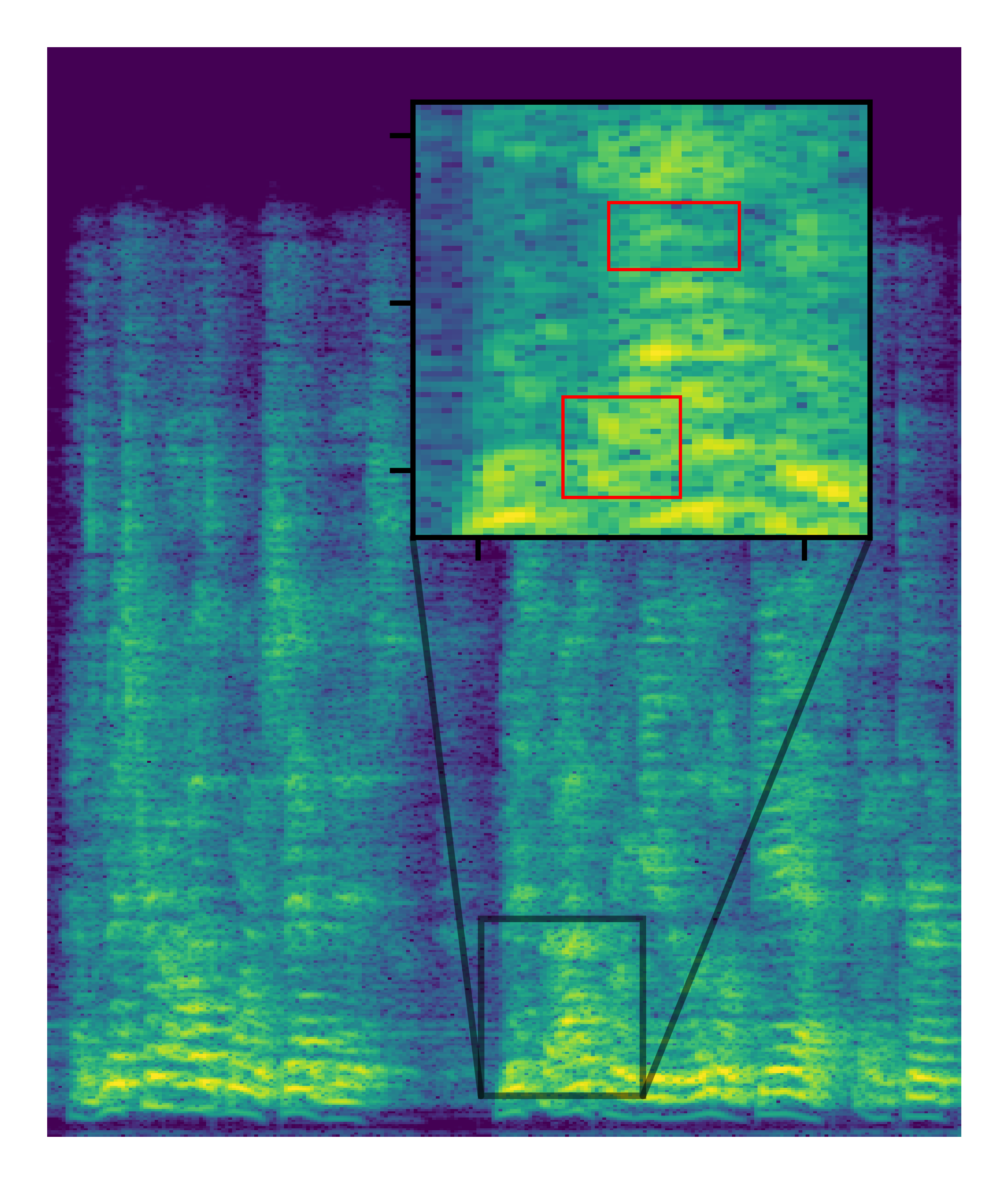}
         \caption{Reference signal}
    \label{fig:Reference}
     \end{subfigure}
     \begin{subfigure}[t]{0.23\textwidth}
         \centering
         \includegraphics[width=0.95\textwidth]{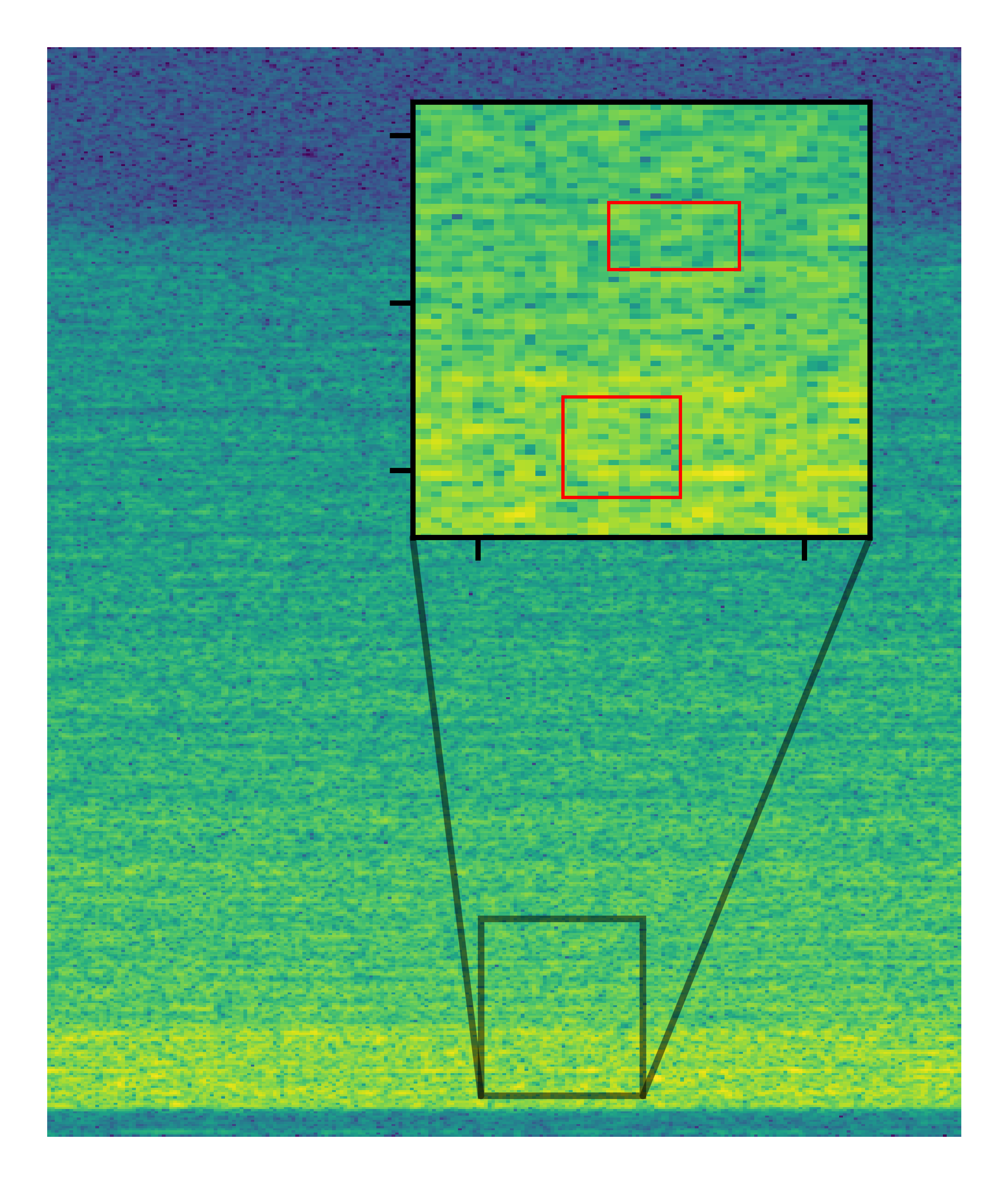}
         \caption{Noisy signal}
         \label{fig:Noisy}
     \end{subfigure}
     \newline
    \begin{subfigure}[b]{0.23\textwidth}
      \centering
       \vspace{0.4cm}
       \includegraphics[width=0.95\textwidth]{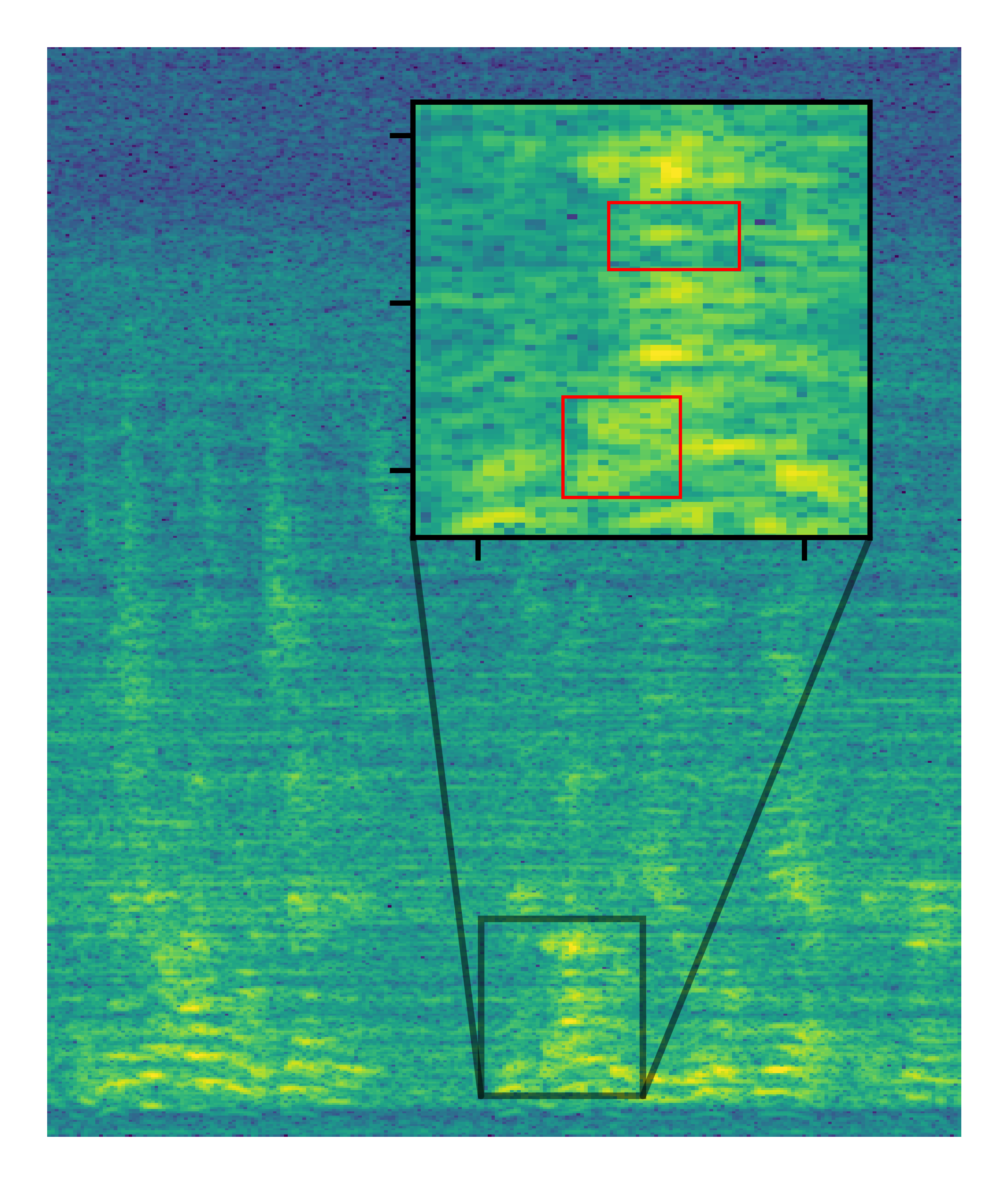}
       \caption{GEVD output}
       \label{fig:GEVD}
     \end{subfigure}
     \begin{subfigure}[b]{0.23\textwidth}
         \centering
         \includegraphics[width=0.95\textwidth]{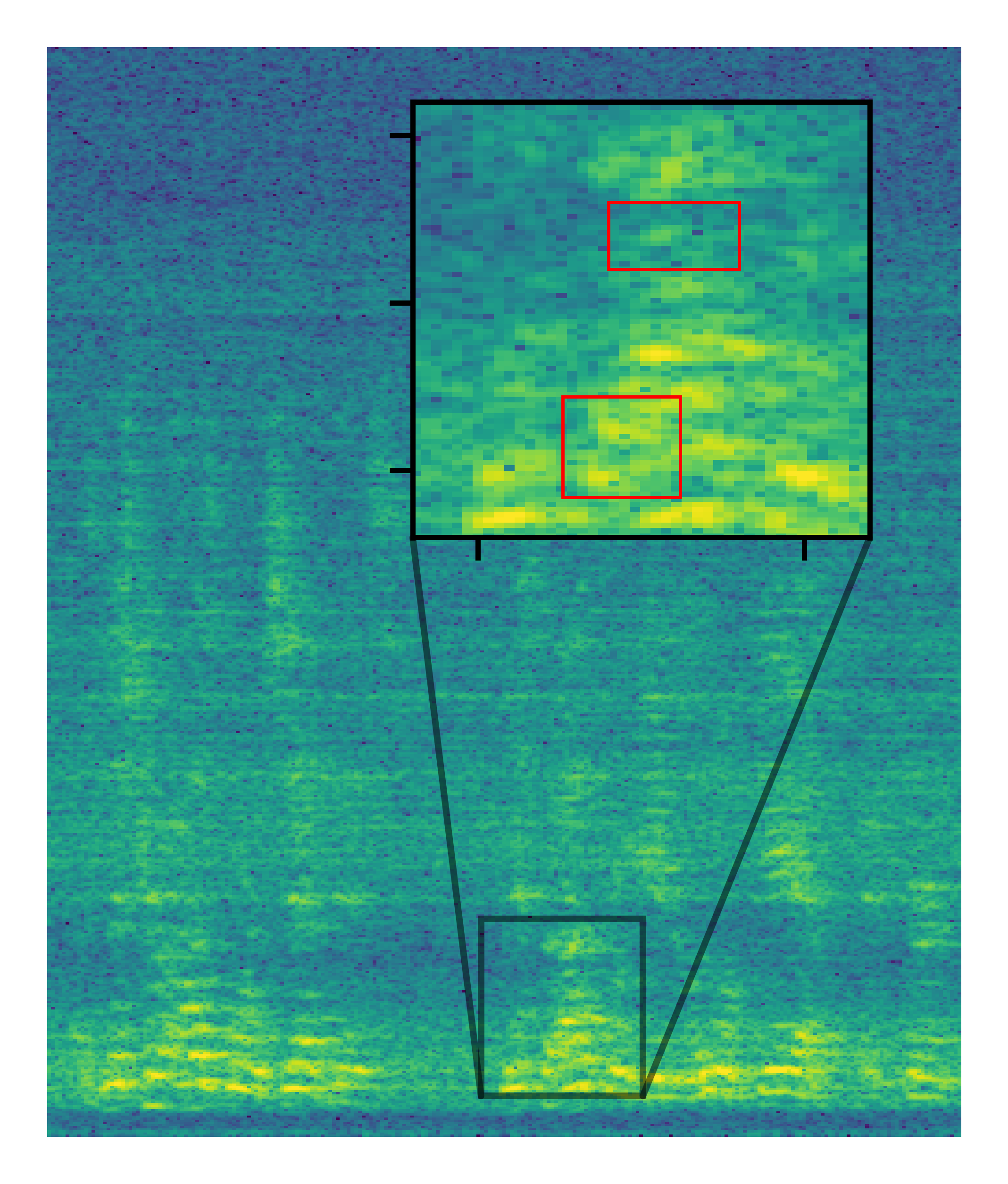}
         \caption{peerRTF output}
         \label{fig:peerRTF}
     \end{subfigure}
       \setlength{\abovecaptionskip}{18pt}
        \setlength{\belowcaptionskip}{-6pt}
        \caption{Sonograms: $\textrm{SNR}_{\textrm{in}}=-10 $~dB and $T_{60}=600$~ms.}
 
        \label{fig:sonograms}
\end{figure}

These advantages are also subjectively demonstrated in Fig.~\ref{fig:sonograms} by sonogram assessment for a randomly chosen example from the test set at $\textrm{SNR}_{\textrm{in}}=-10 $~dB and $T_{60}=600$[ms]. 
We compare the reference signal (the target), the noisy signal, \ac{RTF}-based \ac{MVDR} beamformer output, and our proposed peerRTF beamformer output. We also provide a zoom-in sub-figure to assess the fine details. 
When comparing the beamformer outputs to the reference signal, it is evident that the peerRTF output more closely matches the reference signal than the \ac{GEVD} output. For instance, in the upper rectangle, we can observe a strong frequency bin at the \ac{GEVD} output, which does not appear in either the reference signal or the peerRTF output. In the lower rectangle, there is a small speech gap that is present in both the reference signal and the peerRTF output but missing in the \ac{GEVD} output. Additionally, the peerRTF sonogram exhibits clearer speech patterns and less noise compared to the \ac{GEVD} output, suggesting fewer artifacts and better noise suppression.
%
Sound samples are available on our project page.\footnote{Project Page: \url{https://peerrtf.github.io/}
}

\textcolor{black}{
\subsection{Evaluation with Real-World Noise Types}
To further evaluate the robustness of our method in realistic scenarios, we conducted additional experiments incorporating various environmental noise types that were not included in the training data. These experiments focused on moderate reverberation conditions ($T_{60} = 300$ ms) and tested the method's performance against office, car, and factory noise from the NOISEX-92 database \cite{varga1993assessment} at a challenging SNR level of -10 dB.
The results, summarized in Table~\ref{table:realistic_noise}, indicate that while the performance exhibits a slight degradation when exposed to these previously unseen noise types, the proposed method consistently outperforms baseline approaches across most metrics, especially in low SNR conditions. This resilience can be attributed to the method's reliance on spatial information rather than the spectral characteristics of the noise. Such robustness against diverse environmental noise underscores the method's practical applicability and its strong potential for real-world deployment.
\begin{table*}[htbp]
\caption{Results for various noise types.}
\begin{center}
\resizebox{1.2\columnwidth}{!}{
\begin{tabular}{@{}llccccc@{}}
\toprule
Noise Type & Model & STOI & ESTOI &  SISDR & P808 MOS & SNR\\
\midrule
\multirow{6}{*}{Car Noise} & Unprocessed & 50.81 & 27.62 & -10 & 2.44 & -10 \\
& Reference & - & - & - & 3.3 & - \\
& Oracle & 87.24 & 74.24 & 4.2 & 3 & 20.78 \\
\cmidrule(l){2-7}
& GEVD & 83 & 67.51 & 1.36 & 2.83 & 16.65 \\
& peerRTF & \textbf{86.71} & \textbf{73.74} & \textbf{4.19} & \textbf{3} & 19.83 \\
& CoG & 78.85 & 63.29 & -4.55 & 2.92 & \textbf{21.28} \\
& meanGrid  & 83.59 & 69.33 & 1.29 & 2.82 & 18.36 \\  
\midrule
\multirow{6}{*}{Factory Noise} 
& Unprocessed & 32.35 & 11.53 & -10.05 & 2.48 & -10.00 \\
& Reference & - & - & - & 3.31 & - \\
& Oracle & 85.52 & 71.97 & 3.97 & 2.85 & 23.87 \\
\cmidrule(l){2-7}
& GEVD & 78.32 & 62.08 & -1.85 & 2.80 & 21.74 \\
& peerRTF & \textbf{84.32} & \textbf{70.71} & \textbf{2.8} & \textbf{2.83} & \textbf{24.46} \\
& CoG & 77.00 & 60.79 & -6.64 & 2.72 & 23.23 \\
& meanGrid & 81.81 & 67.46 & 0.29 & 2.82 & 23.51 \\
\midrule
\multirow{6}{*}{Office Noise}  & Unprocessed & 41.99 & 18.99 & -10.07 & 2.65 & -10 \\
& Reference & - & - & - & 3.31 & - \\
& Oracle & 88.15 & 75.66 & 4.67 & 2.93 & 26.17 \\
\cmidrule(l){2-7}
& GEVD & 83.66 & 68.97 & 0.92 & 2.86 & 24.41 \\
& peerRTF & \textbf{87.52} & \textbf{74.88} & \textbf{4.24} & \textbf{2.89} & \textbf{26.49} \\
& CoG & 80.78 & 65.54 & -2.91 & 2.78 & 25.30 \\
& meanGrid & 84.86 & 71.30 & 1.39 & 2.88 & 25.51 \\
\midrule
\multirow{6}{*}{Pink Noise} & Unprocessed & 33.4 & 10.51 & -10 & 2.19 & -10 \\
& Reference & - & - & - & 3.32 & - \\
& Oracle & 87.29 & 74.21 & 4.96 & 3.32 & 24.96 \\
\cmidrule(l){2-7}
& GEVD & 80.57 & -0.77 & -11.76 &2.71 & 21.89 \\
& peerRTF & \textbf{86.17} & \textbf{73.3} & \textbf{4.2} & \textbf{2.82} & \textbf{24.96} \\
& CoG & 79.42 & 63.9 & -3.87 & 2.7 & 23.71 \\
& meanGrid & 84.12 & 70.46 & 1.5 & 2.75 & 24.2 \\
\bottomrule
\end{tabular}}
\end{center}
\label{table:realistic_noise}
\end{table*}
}
\subsection{Alternative Graph Schemes and Loss Functions}
In this section, we examine alternative graph schemes and various loss functions.

First, we compare different graph schemes to demonstrate how the neighboring nodes affect performance. Specifically, we aim to highlight \textcolor{black}{three approaches: 1) processing nodes in isolation (i.e., graphs without edges), 2) using simple averaging of neighboring nodes, and 3) our proposed peerRTF method of learning from neighbor relationships. The comparison with the first scheme, namely the disconnected graph, addresses a provocative hypothesis: is feature enhancement primarily due to the network's power rather than neighbor information? The comparison with the second scheme, namely the neighbor-averaging scheme, provides a baseline that leverages neighboring information without neural network processing, helping determine whether sophisticated neighbor processing genuinely improves performance or if more straightforward approaches suffice.
Our evaluation focuses on challenging acoustic conditions, specifically for a reverberation time of $T_{60}=600$[ms] and a \ac{SNR} of $-10$~dB. These conditions are particularly unfavorable for the vanilla \ac{GEVD}-based estimator. The three node-sharing schemes are compared in Table~\ref{table:neighbors importance}. All compared structures were trained under identical conditions for fair comparison.
}

\textcolor{black}{The results show that the disconnected  graph scheme improves upon \ac{GEVD} in terms of \ac{STOI} measure and \ac{ESTOI}. However, both \ac{GEVD} and mean neighbors scheme show improved performance in terms of \ac{SI-SDR} and \ac{SNR}. The mean-neighbors approach demonstrates consistent improvement over disconnected graphs across all metrics, confirming the value of neighbor information. The proposed peerRTF approach achieves the best performance between all data sharing structures across all metrics, emphasizing the importance of sophisticated neighbor relationship processing through the \ac{GCN} architecture.}
\begin{table}[htbp]
\setlength{\belowcaptionskip}{1pt}
\caption{Comparison between three node-sharing schemes: proposed method (peerRTF), disconnected graph, and mean neighbors-\ac{RTF}.}
\begin{center}
\begin{tabular}{@{}lccccc@{}}
\toprule

Model & STOI & ESTOI &  SISDR & P808 MOS & SNR\\
\midrule

Unprocessed & 23.85 & 9.32 & -10.2 & 2.22 & -10 \\

Reference & - & - & - & 2.94 & - \\

Oracle & 72.07 & 55.95 & 0.57 & 2.53 &  16.83 \\
\midrule

GEVD &  66.52 &  49.5 &  -3.33 &  2.52 & 14.21 \\


peerRTF & \textbf{71.63} & \textbf{55.53} & \textbf{0.46} & \textbf{2.62} &  \textbf{17.3} \\

Self-\acp{RTF} & 68.06 & 51.54 & -6.2 & 2.48 &  12.29 \\

mean neighbors-\acp{RTF} & 68.31 & 51.89 & -3.98 & 2.5 &  16.24 \\
\bottomrule
\end{tabular}
\end{center}
\label{table:neighbors importance}
\vspace{.3cm}
\end{table}

Next, in Table~\ref{table:objective compare}, we examine the impact of the different objective functions. 
\textcolor{black}{Analyzing the table reveals that each objective function favors a different quality measure. Since the differences are not too large, we ultimately selected \ac{SI-SDR} \RNum{2}, as it combines end-to-end audio enhancement with direct \ac{RTF} optimization.}
%
%
\begin{table}[htbp]
\setlength{\belowcaptionskip}{-4pt}
\setlength{\abovecaptionskip}{0pt}
\caption{Comparison of different objective functions for peerRTF training. Results shown for T60 = 600[ms] and input SNR = -10~[dB].}
\begin{center}
\begin{tabular}{@{}lccccc@{}}
\toprule

Model & STOI & ESTOI &  SISDR & P808 MOS & SNR\\
\midrule

Unprocessed & 23.85 & 9.32 & -10.2 & 2.22 & -10 \\

Reference & - & - & - & 2.94 & - \\

Oracle & 72.07 & 55.95 & 0.57 & 2.53 &  16.83 \\
\midrule

GEVD &  66.52 &  49.5 &  -3.33 &  2.52 & 14.21 \\

MP & 70.23 & 54.21 & -1.77 & 2.52 &  15.65 \\

\ac{SI-SDR} \RNum{1} & 72.52 & 57.21 &  \textbf{2.34} & 2.57 &  \textbf{17.96} \\

\ac{SI-SDR} \RNum{2} & 71.63 & 55.53 & 0.46 & \textbf{2.62} &  17.3 \\

SBF & 71.32 & 55.53 & -0.5 & 2.52 &  15.5 \\

STOI & \textbf{72.87} & \textbf{57.32} & -2.23 & 2.51 &  15.12 \\

\bottomrule
\end{tabular}
\end{center}
\label{table:objective compare}
\vspace{-.4cm}
\end{table}

\textcolor{black}{
\section{Additional Simulation Study: Low Grid Resolution}
\label{Additional study}
\subsection{Simulation Setup}
}

\textcolor{black}{In this section, we present an additional simulation study to evaluate the performance of the proposed method with a different array-source constellation, particularly when the number of graph nodes is significantly reduced. This study utilized a publicly available \ac{RIR} generator tool\footnote{\url{https://github.com/ehabets/RIR-Generator}} based on the image method \cite{allen1979image} to synthesize \acp{RIR}.
}
\textcolor{black}{The simulation setup retained the room dimensions and microphone array configuration used in the MiraGe experiments. For this study, we considered only a single reverberation time of $T_{60} = 300$~ms.
}

\textcolor{black}{The source locations were arranged along a semicircular arc with a radius of 1.5~m around the microphone array. For graph construction during the training stage, we used a grid of 180 positions with $1^\circ$ spacing, spanning the range from $0^\circ$ to $180^\circ$. For the test and validation sets, sources were positioned at $5^\circ$ intervals along the same arc (36 positions), with random radial perturbations of $\pm 10$cm sampled uniformly around the nominal radius of 1.5m. The validation and test sets were strictly non-overlapping.
}
\textcolor{black}{To enrich the training dataset, we repeated the training and validation simulations with multiple microphone array positions and orientations within the acoustic enclosure.
}

\textcolor{black}{
This simulation setup differs from our MIRaGe dataset experiments in two key aspects. First, instead of having one fixed graph constructed from all positions in the room, as in MIRaGe, we construct a separate graph for each arc position of the microphone array to ensure sufficient data points for effective training. Second, while MIRaGe used a dense grid of measured positions, this simulation uses a sparser arc-based configuration. For each graph, we use finer angular resolution ($1^\circ$ intervals), while testing is performed at coarser $5^\circ$ increments to better represent real-life scenarios.
}
%
\textcolor{black}{
\begin{figure*}[htbp]
     \centering
    \begin{subfigure}[b]{0.31\textwidth}
         \centering         \includegraphics[width=\textwidth]{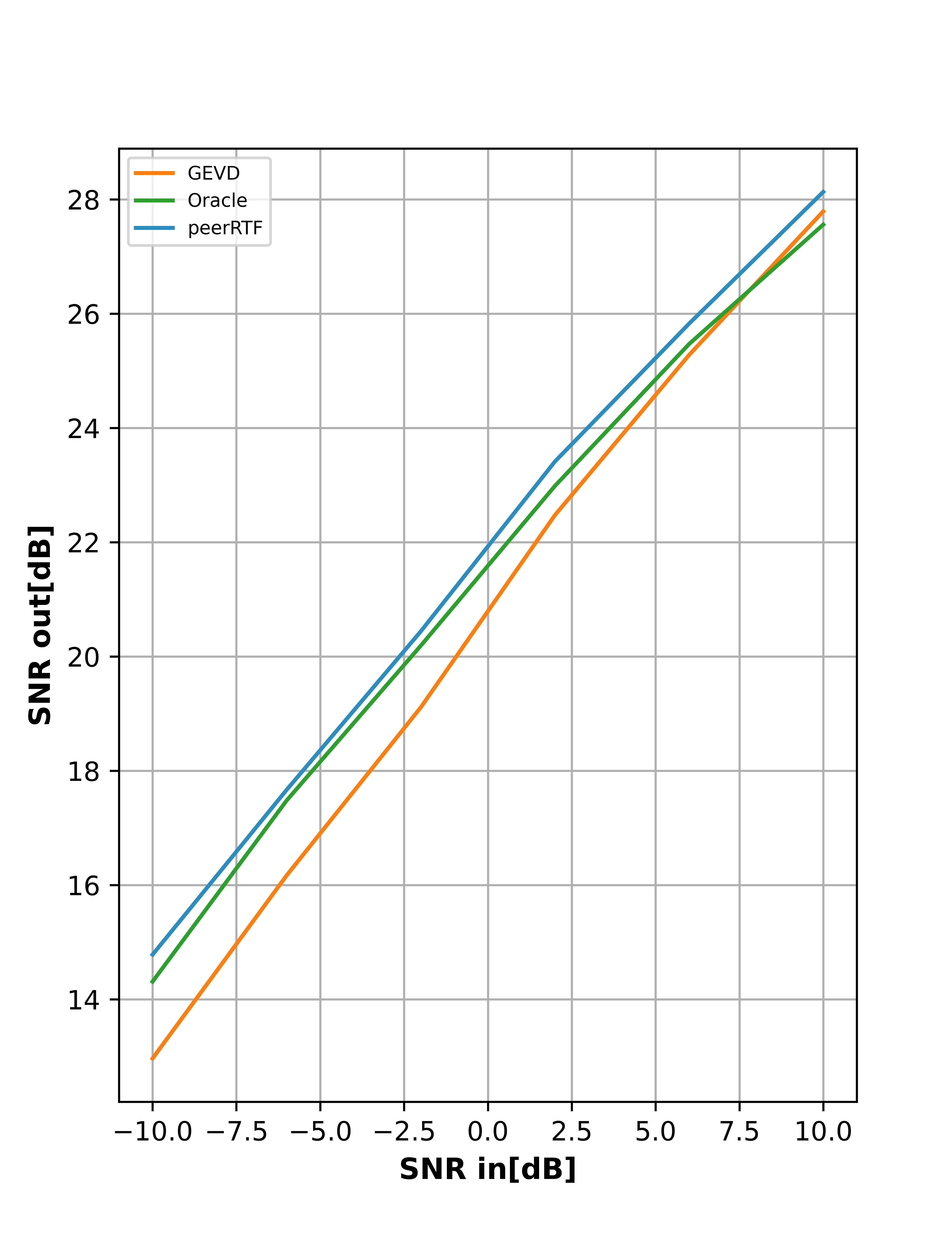}        
    \end{subfigure}
    \begin{subfigure}[b]{0.31\textwidth}
         \centering         \includegraphics[width=\textwidth]{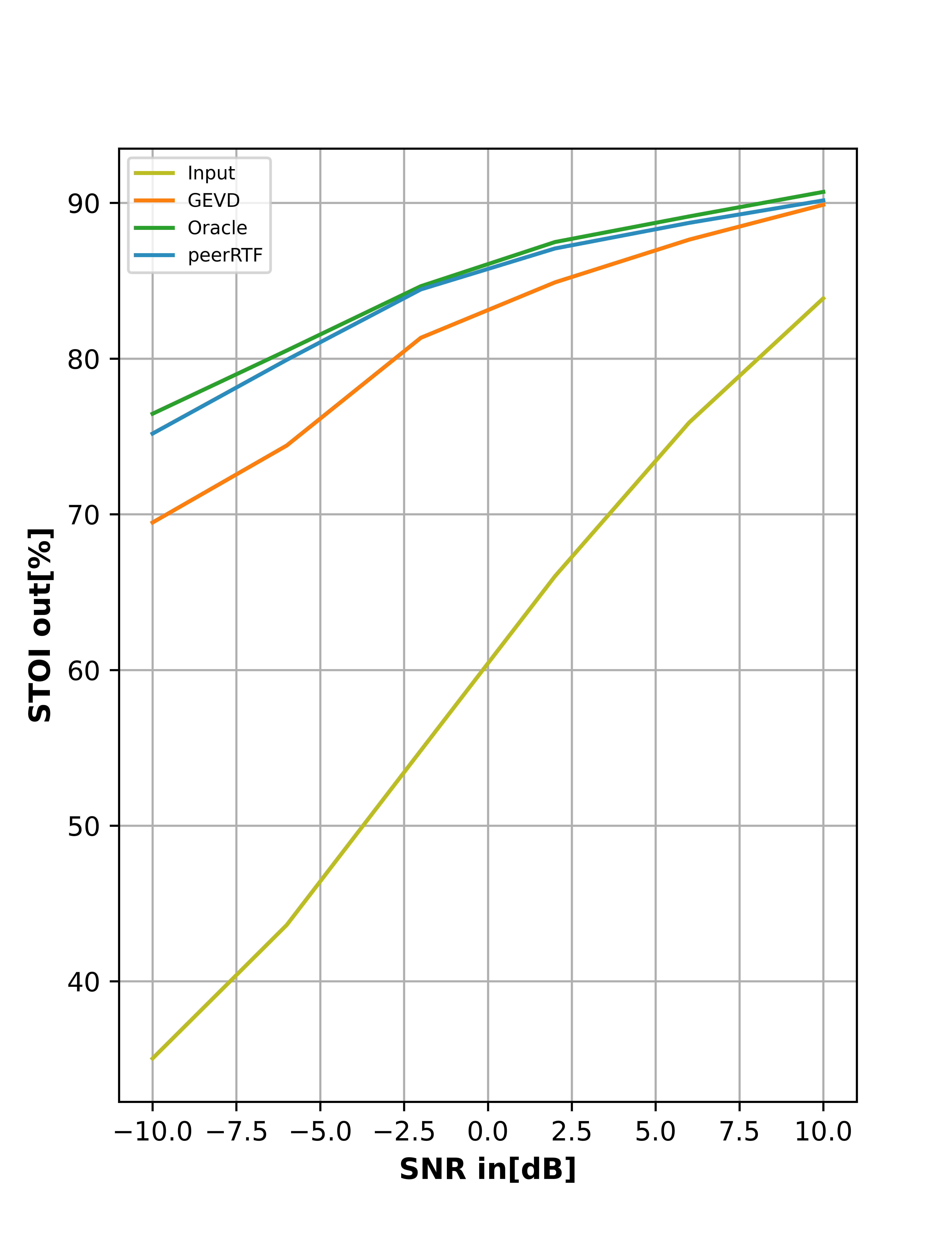}         
     \end{subfigure}
    \begin{subfigure}[b]{0.31\textwidth}
         \centering         \includegraphics[width=\textwidth]{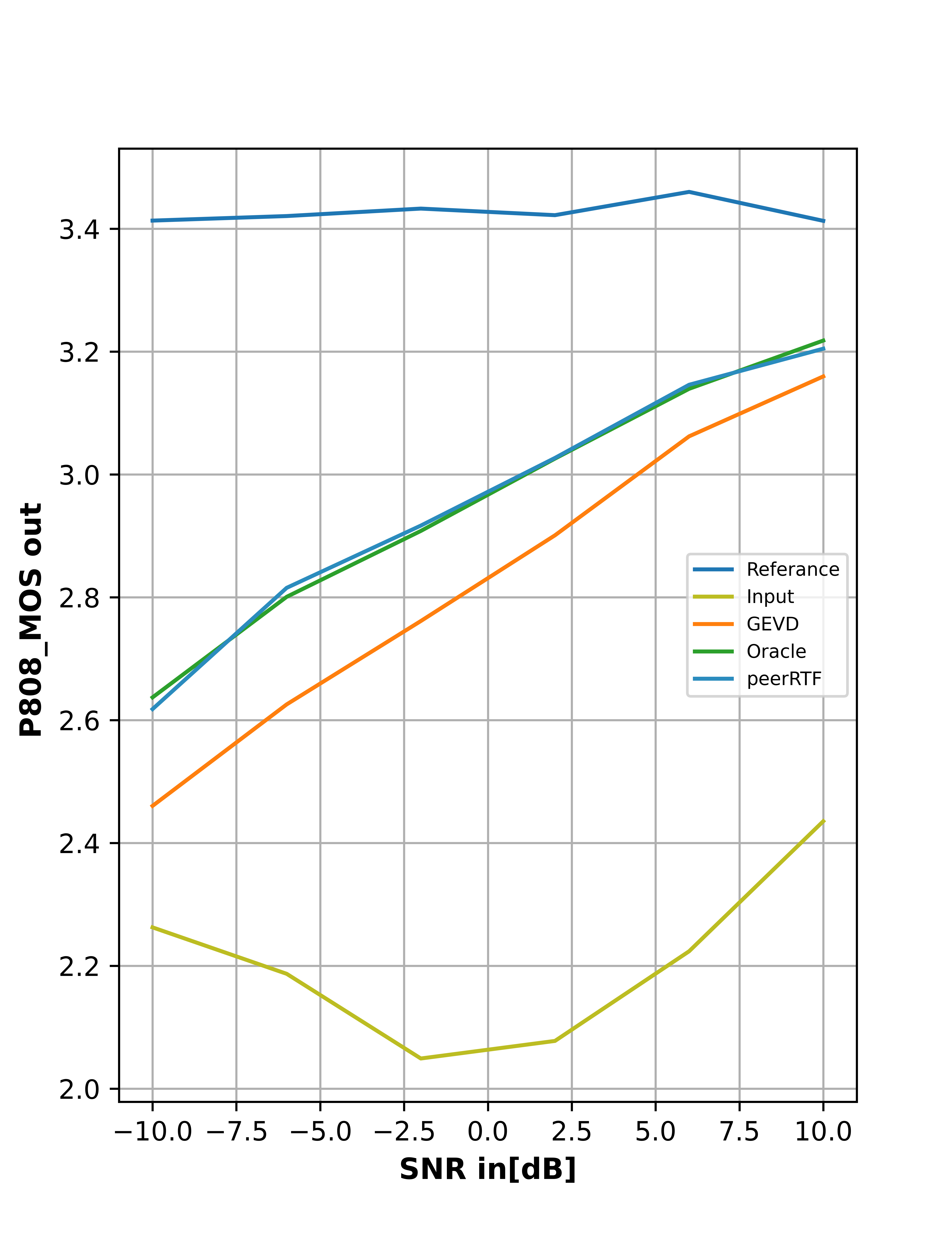}
     \end{subfigure}
\caption{Results for simulated data (semicircular arc) with $T_{60}=300$[ms].}
        \label{fig:results300simulated}
\end{figure*}
}
\textcolor{black}{
\subsection{Results Overview}
This simulation study provides further insights into the performance of the proposed peerRTF method when applied to a different dataset and under reduced spatial resolution in the graph construction. The analysis was conducted using three standard metrics: SNR$_{\text{out}}$ for noise reduction, STOI for speech intelligibility, and DNSMOS for speech quality evaluation.
}

\textcolor{black}{As shown in Fig.~\ref{fig:results300simulated}, the results from this study closely align with the patterns observed in our previous experiments. The proposed method demonstrated consistent performance across various microphone array orientations and source positions, even when using a sparser \ac{RIR} grid. While some performance variations were expected in more challenging scenarios, the overall results suggest that the method effectively adapts to typical acoustic environments.
It is important to note that the source-array constellation remains fixed between the training and test stages, with perturbations to the source distance from the array ($\pm 10$~cm in our experiments). Despite these controlled perturbations, the study confirms the applicability of the proposed method under diverse conditions.
}

%

\section{Conclusion} \label{Conclusion}
In this paper, we have presented a novel \ac{RTF} identification method that relies on learning the \ac{RTF} manifold using a \ac{GCN} to infer a robust estimation of the \ac{RTF} in a noisy and reverberant environment. This approach aims at a robust implementation of acoustic beamforming by utilizing spatial information through the application of \acp{GCN} to this domain. To the best of our knowledge, this is the first time \acp{GCN} have been employed for robust \ac{RTF} estimation, offering a unique way to capture and leverage the complex spatial relationships within the \ac{RTF} manifold.
By utilizing \acp{GCN}, our method explores a different approach to learning-based acoustic processing. It aims to account for the interconnected nature of spatial acoustic information, potentially offering improved robustness in \ac{RTF} estimation under challenging acoustic conditions.
The results presented here, using both simulated and real-life \acp{RIR}, demonstrate the advantages of directly applying a learning algorithm to a graph representing the manifold. This approach is superior to learning a projection of the high-dimensional graph data into Euclidean space, which involves flattening the manifold and performing operations within that space.

\textcolor{black}{Further reducing the number of grid points remains a challenge. Moreover, as with other \ac{ML}-based methods in acoustic signal processing, the ability to generalize from training data captured with a specific array-source constellation to other constellations, or even to a different acoustic enclosure, has yet to be thoroughly explored.
}

\textcolor{black}{In the broader context of deep learning approaches to beamforming, multiple approaches exist to incorporate DNNs into spatial filtering, including mask estimation for MVDR beamforming or direct end-to-end estimation of desired sources using multiple inputs. Our contribution focuses specifically on methods that preserve spatial characteristics, as these can be better analyzed and explained.}

There remain several opportunities to enhance further the \ac{GCN} model and its robustness. Future work could focus on improving the model architecture to achieve better performance by refining the graph structure. For example, exploring more advanced methods for selecting neighbors and defining edges could yield significant benefits. Additionally, the current work has not yet evaluated the model in scenarios involving multiple sources, where accurate \ac{RTF} estimation becomes even more critical. Addressing these challenges would further advance the applicability and effectiveness of the proposed method.

\appendix[Manifold learning \& Graph neural networks]
\label{sec:GCN}

\subsection{Graphs in manifold learning} 

In \acf{ML} problems, we aim to infer a low-dimensional representation of complex, high-dimensional data. Many \ac{ML} algorithms follow a common blueprint for this process. First, they construct a neighborhood graph to capture relationships among data points, which serves as a basis for representing the original data structure. Then, they compute a low-dimensional representation (embedding) of the data, preserving a specific aspect of the original manifold structure. For instance, locally linear embedding \cite{roweis2000nonlinear}, Isomap \cite{tenenbaum2000global}, and Laplacian eigenmaps \cite{belkin2001laplacian} each use different techniques to achieve this. 

\textcolor{black}{Unlike graph-based methods, \acp{VAE} \cite{kingma2013auto} fit into this framework by learning a probabilistic latent space directly, with the encoder output introducing a distributional structure to the embedding. Extensions such as conditional \acp{VAE} \cite{sohn2015learning} further refine this embedding by incorporating conditional information to control the learned representation. Some adversarial autoencoders \cite{makhzani2015adversarial} combine adversarial training to encourage specific forms of structure in the latent space, aiming to produce a more tractable latent space by `flattening' the non-Euclidean structure of the original manifold. 
}

\textcolor{black}{Once inferred, the low-dimensional manifold representation, whether obtained from graph analysis or \ac{VAE} and its variants, can be used in task-specific applications, such as classification, clustering, or regression.}

\textcolor{black}{In \cite{svoboda2019peernets}, the relationship between graph structure, \ac{ML}, and \ac{GNN} is established, demonstrating how the graph structure contributes to the model's accuracy and how \acp{GNN} can effectively leverage this structure. In the next subsections, we explore a particular instance of \acp{GNN}, namely \acp{GCN}.
}
%


\subsection{Graph Convolution Networks} 


\textcolor{black}{In this section, we introduce \acp{GCN}. In the next section, we focus on spatial implementations used in our method.}

\textcolor{black}{A graph $\mathcal{G}$ consists of a node feature matrix $\vect{V} \in \mathbb{R}^{N \times d}$ containing the features of $N$ nodes, and an adjacency matrix $\vect{A} \in \mathbb{R}^{N \times N}$ representing connections between nodes. In our implementation, we use binary connections where $\mathbf{A}_{i,j} = 1$ if node $j$ belongs to the neighborhood of node $i$ (denoted $j \in \mathcal{N}(i)$), and 0 otherwise.}

\textcolor{black}{\acp{GNN} extend conventional neural networks to process graph-structured data by iteratively propagating information through nodes and edges. A key variant is the \ac{GCN}, which, similar to \acp{CNN}, employs shared weights for efficient learning. This is achieved through message passing, where each node aggregates information from its neighbors, enabling the network to capture and exploit the inherent structure encoded in the graph.}

\textcolor{black}{Current \acp{GCN} algorithms can be categorized into spectral-based and spatial-based approaches. Spectral-based methods rely on graph spectral theory, while spatial-based methods operate directly on node neighborhoods. We focus on spatial-based \acp{GCN} \cite{atwood2016diffusion, velivckovic2018graph, wang2019dynamic}  as they are well-suited for node-specific tasks like ours by operating locally on each node without requiring global graph information.}

\subsection{Spatial GCN}
\textcolor{black}{Spatial \acp{GCN} extend the concept of convolution from regular grid structures, like images, to irregular graph structures. In traditional \acp{CNN}, each pixel aggregates information from its neighboring pixels through weighted averaging. Similarly, in \acp{GCN}, each node aggregates information from its neighboring nodes, but through a more flexible mechanism.  A schematic comparison between 2D convolution and graph convolution is depicted in Fig.~\ref{fig:conv_vs_gcn}.
\usetikzlibrary{positioning}

\tikzset{darkstyle/.style={circle,draw,fill=gray!40}}
\tikzset{centerstyle/.style={circle,draw,fill=red!40}}
\begin{figure}[t]
\centering
\begin{tikzpicture}
	
	 \scalebox{0.62}{
	\begin{scope}[xshift=-0.6cm,yshift=-3cm]
            	\foreach \i in {1,...,6}
	{
		\pgfmathtruncatemacro{\y}{(\i - 1) / 5}
		\pgfmathtruncatemacro{\x}{\i - 5 * \y}
		\pgfmathtruncatemacro{\label}{\x + 5 * (4 - \y)}
		
		\node[darkstyle,minimum size=20] (\label) at (1.5*\x,1.5*\y)
		{\label};
	}
            	\foreach \i in {8,...,25}
	{
		\pgfmathtruncatemacro{\y}{(\i - 1) / 5}
		\pgfmathtruncatemacro{\x}{\i - 5 * \y}
		\pgfmathtruncatemacro{\label}{\x + 5 * (4 - \y)}
		
		\node[darkstyle,minimum size=20] (\label) at (1.5*\x,1.5*\y)
		{\label};
	}

	\pgfmathtruncatemacro{\y}{(7 - 1) / 5}
	\pgfmathtruncatemacro{\x}{7 - 5 * \y}
	\pgfmathtruncatemacro{\label}{\x + 5 * (4 - \y)}
		
	\node[centerstyle,minimum size=20] (\label) at (1.5*\x,1.5*\y)
		{\label};
	
	
	\foreach \x in {1,...,4}
	\foreach \y in {0,...,4}
	{
		\pgfmathtruncatemacro{\cur}{\x + 5* \y}
		\pgfmathtruncatemacro{\next}{\cur + 1}
		\draw (\cur) -- (\next);
	}
	
	\foreach \x in {1,...,4}
	\foreach \y in {0,...,3}
	{
		\pgfmathtruncatemacro{\cur}{\x + 5* \y}
		\pgfmathtruncatemacro{\next}{\cur + 1}
		\pgfmathtruncatemacro{\botri}{\cur+6}
		\pgfmathtruncatemacro{\botle}{\cur + 5}
		\draw (\cur) -- (\botri);
		\draw (\next) -- (\botle);
	}
	
	\foreach \start in {1,...,20}
	{
		\pgfmathtruncatemacro{\down}{\start+5}
		\draw (\start) -- (\down);  
	}
	
		\draw[->, very thick] (11)[red] -- (17);
		\draw[->, very thick] (12)[red] -- (17);
		\draw[->, very thick] (13)[red] -- (17);
		\draw[->, very thick] (16)[red] -- (17);
		\draw[->, very thick] (18)[red] -- (17);
		\draw[->, very thick] (21)[red] -- (17);
		\draw[->, very thick] (22)[red] -- (17);
		\draw[->, very thick] (23)[red] -- (17);
	
    \draw[blue,thick,dashed,rounded corners=15pt,fill opacity=0.15,fill=blue!]
  (,-0.5) rectangle ++(4,4);
  
	\end{scope}}
	 \scalebox{0.78}{
	\begin{scope}[xshift=9.5cm]
        	\tikzstyle{every node}=[draw,shape=circle,inner sep=1,minimum size=1.5em];

    \node[centerstyle] (v0) at (0:0) [fill] {};
	\node[darkstyle] (v1) at ( 20:1.7) [fill] {};
	\node[darkstyle] (v2) at ( 72:2) [fill] {};
	\node[darkstyle] (v3) at (2*72:1.5) [fill] {};
	\node[darkstyle] (v4) at (3*72:2) [fill] {};
	\node[darkstyle] (v5) at (4*72:1) [fill] {};
	\node[darkstyle] (v6) at (100:3) [fill] {};
	\node[darkstyle] (v7) at (130:3) [fill] {};
	\node[darkstyle] (v8) at (120:4) [fill] {};
	\node[darkstyle] (v9) at (250:3) [fill] {};
    \node[darkstyle] (v10) at (150:3.2) [fill] {};
	\draw[<-, very thick] (v0)[red] -- (v1);
	\draw[<-, very thick] (v0)[red] -- (v2);
	\draw[<-, very thick] (v0)[red] -- (v3);
	\draw[<-, very thick] (v0)[red] -- (v4);
	\draw[<-, very thick] (v0)[red] -- (v5);
	\draw (v7) -- (v3);
    \draw (v8) -- (v7);
    \draw (v6) -- (v3);
    \draw (v6) -- (v2);
    \draw (v6) -- (v8);
    \draw (v9) -- (v4);
    \draw (v2) -- (v7);
    \draw (v9) -- (v5);
    \draw (v10) -- (v7);
    \draw (v10) -- (v8);
    \filldraw[fill opacity=0.15,fill=blue!,blue,thick,dashed] (0,0) circle (2.35cm);
    
	\end{scope}}

\end{tikzpicture}
\caption{Comparison of 2D convolution and graph convolution. 
\textbf{Left:} In conventional 2D convolution on Euclidean data, such as an image, the central pixel (shown in red) is computed as a weighted average of itself and its neighboring pixels based on the kernel size. The ordered grid structure provides a consistent spatial arrangement.
\textbf{Right:} In spatial graph convolution, the node representation is computed by aggregating features from neighboring nodes without relying on any fixed spatial ordering or grid structure. (Inspired by \cite{cao2020comprehensive}).}
\label{fig:conv_vs_gcn}

\end{figure} 
Unlike \acp{CNN}, which use scalar multiplications based on fixed spatial positions, \acp{GCN} process node relationships through small neural networks (depicted as red arrows in Fig.~\ref{fig:conv_vs_gcn}). These networks can be simple nonlinear transformations or more complex \acp{MLP}. A key feature of \acp{GCN} is their permutation invariance - the output does not depend on the ordering of neighboring nodes. This is achieved by sharing weights uniformly across all node relationships throughout the graph.}

%
\textcolor{black}{A spatial \ac{GCN} processes information through multiple graph convolution layers. Each layer: 1) aggregates features from neighboring nodes and 2) applies nonlinear transformations via \acp{MLP}  to the aggregated features. The network's depth determines the extent of information propagation, with deeper networks accessing higher-order neighbor relationships \cite{alon2020bottleneck}. In our implementation, we limit this to first-order neighbors, with justification provided in Sec.~\ref{Our method}.}


\textcolor{black}{While \acp{GCN}  are commonly used for classification tasks, we extend them to perform regression on high-dimensional, continuous-valued vectors. This approach allows us to learn node representations that capture both local graph structure and the complex relationships between nodes, leading to more accurate predictions for our \ac{RTF} estimation task.}


%






\ifCLASSOPTIONcaptionsoff
  \newpage
\fi



\bibliographystyle{IEEEtran}
\bibliography{IEEEabrv,ref}
%
%








\end{document}